%% file: PSCarXiv.tex
\documentclass[12pt]{article}
\usepackage{graphicx}

\newcommand{\beq}{\begin{equation}}
\newcommand{\eeq}{\end{equation}}
\newcommand{\bea}{\begin{eqnarray}}
\newcommand{\eea}{\end{eqnarray}}

\begin{document}
\begin{center}
{\LARGE Multifractal Analysis of Packed Swiss Cheese Cosmologies}
\vskip .5cm
J.\ R.\ Mureika\\
{\it W.\ M.\ Keck Science Center, The Claremont Colleges, Claremont, CA~91711~USA} \\
Email: jmureika@jsd.claremont.edu \\
\vskip .3cm
C.\ C.\ Dyer\\
{\it Department of Astronomy and Astrophysics, University of Toronto, Toronto, ON~M5S~1A7~Canada}
\vskip .5cm

\vskip 2cm

\end{center}
\noindent
{\footnotesize
{\bf Abstract} \\
The multifractal spectrum of various three-dimensional representations of 
Packed Swiss Cheese cosmologies in open, closed, and flat spaces are measured, 
and it is determined that the curvature of the space does not alter the 
associated fractal structure.  These results are compared to observational data 
and simulated models of large scale galaxy clustering, to assess the viability 
of the PSC as a candidate for such structure formation.  It is found that
the PSC dimension spectra do not match those of observation, and possible 
solutions to this discrepancy are offered, including accounting for potential 
luminosity biasing effects.  Various random and uniform sets are also analyzed 
to provide insight into the meaning of the multifractal spectrum as it relates 
to the observed scaling behaviors.
}

\pagebreak
\section{Do We Live in a Fractal Universe?}
\label{review}
The notion of a hierarchically-structured world is a recurrent
theme in our understanding of Nature \cite{mandel1}.
According to the Cosmological Principle, the Universe must be 
homogeneous and isotropic.  This oft cited-as-fact stipulation is the
basis for the Friedmann-Robertson-Walker (FRW) solutions to Einstein's
Field Equations, from which the expansion dynamics of the Universe are
derived \cite{mtw}.  However, the breakdown of homogeneity (at least on smaller
distance scales in the Universe) is quickly becoming an accepted ideal
in cosmological circles.  That is, while homogeneity requires matter
to scale uniformly in space ({\it i.e.} $D_F = 3$, equal probability 
scaling behavior
in all spatial directions), actual measurements
of the distributions suggest otherwise.

The existence of fractally-clustered matter was further emphasized by
Peebles (see \cite{peebles2}), whose various two-point correlation analyses 
of three-dimensional catalogs yielded the exponent $\gamma \sim 1.7$, the 
co-dimension 
of which was taken to be the fractal dimension $D_F = 3-\gamma \sim 1.3$.
Some successive works confirmed this value, showing a study of the CfA
redshift survey to agree statistically \cite{apj332,apj357}.  However,
these same studies cite model simulations with higher dimensionality,
in particular $D \sim 2$\footnote{This discrepancy ultimately has its roots in
the method of calculation.  The dimension $D=D_F=2$ is the standard fractal
dimension obtained by usual means, while The value $D \sim 1.3$, 
calculated from a two-point correlation function, is actually the $q=2$ 
multifractal dimension $D_2$.  See Section~\ref{obscomp}}.

Over the course of a decade, increasing evidence has been put forth to suggest
that the dimensionality of a wide range of redshift survey catalogs yield
such clustering dimensions.  In 
the comprehensive publication \cite{piet1}, the results
of fractal dimension analysis of numerous galactic catalogs are reviewed,
with the general consensus that each data set reveals a unanimous
$D_F \sim 2$ (see Table~\ref{galaxydims}).  

Since each catalog is
limited in size and spatial extent ({\it i.e.} volume), 
the associated dimensions can only be statistically viable up to some 
effective radius $R_{\rm eff}$ (the radius of the largest sphere one
can inscribe in the associated sample, without surpassing the catalog
boundaries).  
For most galaxy catalogs considered in the literature, 
these range up to 50~$h^{-1}$~Mpc (recall that $h^{-1}$ is the scaling
factor in the Hubble Constant, $H_0 = 100\;h~$km$\;$s$^{-1}$Mpc$^{-1}$
\cite{piet1}).  Pietronero {\it et al.} 
further extend the analysis to include clusters and superclusters,
concluding that the two sets of data represent the same self-similar 
structure at differing scales.  While clusters are more distant, they are
more luminous, and by counting the clusters as single objects, a coarser
estimate of scaling is obtained over larger distances.  This provides
evidence for fractal behavior up to distances 1000~$h^{-1}$~Mpc, with
no suggestion that homogeneity ensues.

Of course, a potential limitation of such estimates is the lack of a definite
third dimension.  While angular spans may be accurately measured (barring
external interference from sources such as gravitational lensing), the
(co-moving) distance to the objects in question must rely on estimates from 
luminosity-distance relations or redshift measurements.  These effects
are introduced primarily by the recessional velocities of the objects, and
in general may be obtained via application of the Hubble redshift
law, whose simplest form for low redshifts is linear,
\beq
r = \frac{cz}{H_0}~,
\label{linhub}
\eeq
and for general redshifts (Mattig's 1958 relation; see {\it e.g.} \cite{piet1})
\beq
r = \frac{c}{H_0}\frac{zq_0+(q_0-1)(\sqrt{2zq_0+1}-1)}{q_0^2\;(1+z)}~,
\label{frwhub}
\eeq
in an expanding Universe with deceleration parameter $q_0$.  For 
$q_0 = 1/2$ ({\it i.e.} flat Universe), this can be shown to reduce to 
\cite{piet1}
\beq
r = 6000 \; \left(1 - \frac{1}{\sqrt{1+z}}\right)~h^{-1}\;{\rm Mpc}
\label{mattig}
\eeq
Typical redshifts for the majority of nearby
catalogs are $z < 0.05$, but deeper surveys such as ESP or LCRS can contain
redshifts of the order $z \sim 0.2$ \cite{eso1,lcrs}.  
Note that there is no correction for peculiar velocities in the associated 
figures of Table~\ref{galaxydims} \cite{piet1}.

The authors of \cite{piet1} conclude that use of the 
Euclidean Hubble relation (\ref{linhub})
in catalogs where the redshifts are high instead of (\ref{frwhub}) or
(\ref{mattig})
does not affect the estimated dimensionality, and thus the result
$D_F \sim 2$ is stable to such variations.  This claim is further elucidated
in reference~\cite{aa367}, who note that use of the Euclidean 
Hubble distance formula may be exported to distance greater than 600~Mpc 
with no consideration of relativistic curvature effects\footnote{
In fact, this provides alternative support for the notion of ``deformation 
independence'' of the fractal dimension with respect to the embedding
manifold, discussed in \cite{mythesis}.}.

Before proceeding, however, it should be emphasized that the aforementioned
$D_F \sim 2$ fractal structure of the local Universe (with no transition
to $D_F = 3$) is a hotly debated subject, and is by no
means to be interpreted as ``fact''.  Rather, it is a movement largely
spearheaded by the authors of \cite{piet1}, whom collectively have published
over 50 related articles in the past 5 years.  In fact, several earlier works 
\cite{peebles2} suggest that while the local structure may be fractal,
the transition to homogeneity is clearly marked at about 5~$h^{-1}$~Mpc.
The general consensus tends to reflect this finding ({\it i.e.} that
at least the {\it local} clustering structure can be described as a fractal), 
although the transition distance varies from catalog to catalog and analysis.
More recently, the authors of reference~\cite{eso} submit that there
is no statistically or physically viable way to obtain a fractal 
scaling behavior for the ESP redshift survey and Abell clusters, 
instead citing a homogeneous $D \sim 3$~scaling, attributing the 
potential $D_F \sim 2$ to various redshift-related distance biases on the
part of Pietronero {\it et al.}, including the substitution of the
Euclidean Hubble law in lieu of the FRW version! A rebuttal to this
publication suggests that the cross-over in question is explicitly
dependent on the cosmological model used for the calculations 
\cite{joyce}.  Clearly, there is much
disagreement in the literature, so one must be careful not to take 
each report at unquestioning face value.

As newer and more comprehensive redshift data becomes available from
such current surveys as the Sloan Digital Sky Survey (SDSS) \cite{sdss} or
the 2dFGRS \cite{twodf} (which are targeting well in excess of 100~000
redshifts, and in the case of SDSS, 1~000~000 galaxies), the 
crux of this debate may
be addressed with more certainty (or, on the other hand, such may serve
to further complicate the issue!).

\section{The Packed Swiss Cheese Cosmological Model}
\label{swiss}
The notion of a fractal Universe defies the Cosmological Principle, which 
demands homogeneity and isotropy at all points.  Locally, however, it is
rather evident that the universe is {\it not} homogeneous.  Any cosmological
model which is used to represent the observed galaxy distribution must 
adhere to this point, lest its power of predictability be diminished.

Some of the earlier references to locally inhomogeneous cosmological models
date back to Einstein and Strauss \cite{originalsc}, 
as well as Schucking in the 1950s \cite{originalsc}, and later
Rees and Sciama in the 1960s \cite{originalsc}.  Dubbed ``Swiss Cheese''
models, these constitute locally inhomogeneous but globally homogeneous
spacetimes which everywhere satisfy Einstein's Field Equations.
Further works studied the effects of multiple hole solutions
vis-a-vis gravitational
lensing effects with and without a Cosmological Constant
\cite{dyerthesis,multihole}, and most recently the notion of
optimally packed, volume-filling hole solutions, dubbed
{\it Packed Swiss Cheese} cosmologies (PSC) \cite{packsc,cheese1}.

A sphere is inscribed in a zero-pressure, 
expanding FRW Universe of spatially-uniform density profile $\rho_0$,
and the mass within is condensed to a smaller sphere of larger
average density $\rho_1 > \rho_0$.  The total mass is conserved within 
the shell.  Outside the inscribed radius, the space is still 
purely FRW, with line element
\bea
ds^2&=& dt^2 - R^2(t) \left[d\omega^2 + S_k^2(\omega) d\Omega^2\right]~, \nonumber \\
 &=& dt^2 - R^2(t) \left[\frac{dr^2}{1-kr^2} + r^2 d\Omega^2\right]~,
\label{frwmetric}
\eea
for $\omega$ the angular radial coordinate, and the co-moving surface
defined by $S_k(\omega)$ is either
$\sin(\omega), \omega, \sinh(\omega)$ for positive, flat, or negative
curvatures spaces respectively (equivalently one can write $S_k(\omega) = 
\sin(\sqrt{k}\omega)/\sqrt{k}$ for $k=+1,0,-1$) 
\cite{mtw,multihole}.  $d\Omega = d\theta^2 + \sin^2(\theta) d\phi^2$ 
is the standard (spherical) solid angle element.  Hence, the CP
is preserved on the exterior.

The metric of the vacuum interior of the hole is the Schwarzschild line element,
\beq
ds^2 = \left(1-\frac{2Gm}{r} -\frac{\Lambda r^2}{3}\right) dt^2 
- \frac{dr^2}{1-2Gm/r - \Lambda r^2/3} - r^2 d\Omega^2~,
\eeq
for general Cosmological Constant $\Lambda$ and interior mass $m$
(subject to appropriate matching conditions at the surface boundary
\cite{dyerthesis}).
If the internal mass density has some spatial extent, this can 
also possess an FRW line element (but not necessarily), although in
the case of the PSC one considers only an interior point mass.
Since their average density profile is unchanged, any number of holes 
may evolve independently of each other, since they have no gravitational
influence on one another (provided the inscribed surfaces do not 
overlap) \cite{multihole}. 

The PSC is conceptually
similar to the classic ``Apollonian Packing'' problem of 
efficiently filling a region with tangential circles of varying sizes.
In the former case, however, it is a three-dimensional packing
problem with spheres in spaces of constant curvature.
A point is chosen at random in a region of FRW space, as before, 
and a sphere of an arbitrarily
large radius $R_0$ is inscribed.  The mass within this sphere is shrunken down
to a smaller radius $R_0'$, increasing the local density 
and creating a density discontinuity at the boundary.  
A second point is chosen in the remaining continuous density region, and
a second sphere is inscribed, subject to the constraint that it be tangential
to the first.  The interior mass is again contracted to a specified radius,
and the process is repeated indefinitely.  As mentioned, the
contracted radius $R' \rightarrow 0$.

Any number of packings and configurations can be obtained by varying the
initial size and placement of the first sphere.  Since there are no
explicit scale constraints on the configurations, a fully recursive PSC
can be formed by inserting within any sphere of one packing the contents
of another.  Such a model hearkens of self-similarity, at the very least
on a statistical level.  The libraries used for this study (see \cite{cheese1}
for details) contain on
average between 30~000 to 90~000 spheres (generally much larger than the size
of the redshift surveys).  Figure~\ref{packgif} shows a two-dimensional
projection of a typical packing library.

There is no discussion herein of effective scales or cutoff radii, as with
the survey catalogs, since the packing libraries are scaled in 
dimensionless units.  Indeed,
since one may recursively pack them at will, the overall physical scales
may be set arbitrarily.

It should be noted that this Swiss Cheese model is different than that of Ribeiro
\cite{ribeiro1,ribeiro2,ribeiro3}, who also predicts a large-scale
fractal structure.  Comparisons with the aforementioned results are
discussed in section~\ref{ribeirowork}.

\subsection{Justification for a Swiss Cheese Cosmology}
\label{pscjustify}

The choice of a Swiss Cheese cosmology may seem at first to be without
clear motivation, but it is in fact one of the most logical and consistent
spacetimes in which to work.  The Swiss Cheese solution is an exact solution 
of the Einstein Field equations at {\it all the length scales involved}. 
This is absolutely necessary to begin consideration of fractal behavior 
since the very essence of the fractal approach is validity over a range of 
length scales.   

The PSC model considered herein is built upon the Swiss Cheese solution, and while this solution at first appears to some to be 
artificial, it covers the essential field regions for an object embedded in a 
background universe. Thus it encompasses a near field, a medium field 
({\it i.e.} the vacuum around the central object) and the far field, which just
becomes the FRW background universe.  The sharpness of the boundaries
have little physical impact.  Note that the sharp boundary of the
Earth does not have any drastic effects when scene in terms of the
gravitational potential (and thus the metric tensor either).

\subsection{Cosmological Principle, GR Style}
\label{grcp}
It is useful to re-formulate the Cosmological Principle in the language
of General Relativity, insofar as the PSC models are concerned.  
Recall that the curvature of a manifold with metric $g_{ab}$ is defined
via second derivatives of the metric, and this information is completely
contained within the Riemann Tensor, 

\beq
R_{abcd} =  g_{ai} [\Gamma^i_{bd,c}
-\Gamma^i_{bc,d} + \Gamma^i_{nc}\Gamma^n_{bd} - \Gamma^i_{nb}\Gamma^n_{cd}]~,
\eeq
(with $\Gamma^i_{jk}$ the associated Christoffel Symbols).  

The Weyl Tensor in $n$ dimensions is \cite{hawkellis}
\beq
C_{abcd} = R_{abcd} + \frac{2}{n-2}\left(g_{a[d}R_{c]b}+g_{b[c}R_{d]a}\right)
+ \frac{2}{(n-1)(n-2)} R\;g_{a[c}g_{d]b}~.
\label{weylten}
\eeq
where $[~]$ denotes the anti-symmetric sum with respect to index permutation.
Since it can be shown that $C^a_{bad} = 0$, the Weyl Tensor
is the trace-free portion of the Riemann Tensor, and
complements the information contained within the Ricci Tensor (trace of
Riemann).  Combined, both objects contain the complete curvature information of
the Riemann tensor, with Ricci representing the local curvature
contributions (via Einsetin's equations, {\it i.e.} as a function of 
the local pressure $p$ and matter density $\rho$ via the energy-momentum 
tensor $T_{ab} = pg_{ab} + (\rho+p)u_au_b$, with $u_a$ the associated 
four-velocity), and Weyl the non-local or external contributions.

A different interpretation is as follows.
For a sphere of radius $r$ inscribed in a density 
field $\rho(r)$, the Weyl
Tensor is a measure of the over/under-density of integrated (or enclosed) mass
$\tilde{m}(r)$ to ``average contained mass'' $\bar{m}(r)$ 
%(see Figure~\ref{weyl}).  
That is, 
$C_{abcd} \propto \tilde{m}(R) - \bar{m}(R)$, where 
\bea
\tilde{m}(R)&=&4\pi\int_0^R \rho(r)\;r^2\:dr~, \nonumber \\
\bar{m}(R)&=&(4\pi/3) R^3 \rho(R)~, 
\eea
and $C_{abcd}$ is evaluated on
some closed surface which encloses the density field $\rho$ (see
reference~\cite{dyerthesis} for a complete derivation).  
If this quantity is non-zero, it signals an imbalance in the mass 
distribution, and hence provides a ``gravitational compass''.  That is, 
by providing a preferred direction, it breaks the isotropy of the space.
If, on the other hand, $C_{aibj}$ vanishes on this surface, then the
space contains no local tidal forces (no compass), 
and the Cosmological Principle is upheld if the
mass distribution is homogeneous.  

Every inscribed sphere in the PSC packings is itself such a surface (see 
Figure~\ref{packgif}), and thus it is always possible
to find a compound surface within the packing on which the Weyl Tensor
vanishes.  The space outside the inscribed boundary is FRW, by design,
but that contained within the sphere is not.  In the PSC model, 
it is assumed to be a vacuum, and thus has vanishing $T_{ab}$
(and thus vanishing Ricci Tensor).

\section{Measuring the Multifractal Spectrum of the PSC Models}
\label{swisscomp}
Effective scaling dimensions are obtained by the (three-dimensional)
box counting method.  Tables~\ref{flat1}-\ref{neg1} 
show the calculated dimension $D_q$,
$q = 0,2$ (box and correlation) and $q \rightarrow \infty$, 
for several packings libraries
of flat, positive, and negative curvatures respectively.
Values of $D_q$ for $q < 0$ are discussed in Section~\ref{negqs}.

Covering cubes of side $d$ ranged roughly over two orders of magnitude, requiring
8 cubes at the largest scales, to about half the population size of
the packings ({\it i.e.} about 1-2 particles per box on average).
Below this limit, the box scales drop below the average interparticle
distance, and the calculation becomes skewed by the finiteness of the
data set.  
Since the libraries are themselves spherical distributions of 
points, 
then the largest box used in the counting is that which optimally fits
within the spherical region, in order to avoid any spatial biasing of 
near-empty boxes in which there are few points ({\it i.e.} at the edges).  
This cubical subset accordingly contains a
reduced fraction of the total library population as listed in the
associated Tables, but still provides for a decent statistical sampling
of the distribution.  Figures~\ref{flatfit}-\ref{negfit} demonstrate the box 
counting regression with associated
confidence levels for the $q=0$, yielding the appropriate slopes.
Note that ``$q \rightarrow \infty$'' cannot be numerically realized,
so the associated values cited herein correspond to 
the values of $D_q$ for $q \geq 60$, at which stage the estimates are
observed to have reached a relatively stable value.

There is a fairly narrow range
in the dimensions for each library, essentially yielding a possible range
of dimensions between $D_0 = 2.5-2.8$, depending on the overall size
of the sample and the choice of fit points, with correlation dimensions
$D_2 = 2.4-2.6$ (see also Figures~\ref{dqlibs1}-\ref{dqlibs3} for the
general form of the $D_q$ spectra).  The results might
suggest that the positively- and negatively-curved spaces yield
slightly higher box dimensions than the flat case, although to within
the cited fit error no definite determination may be made.  In fact,
to the accuracy of the fits, the Tables are virtually indistinguishable.

Note that $D_{q \rightarrow \infty}$ tends to approach a value near
2 in every case.  This result has a simple geometric interpretation, and
in fact provides a definite signature for the PSC mechanism.  Since the
$D_q$ values for large $q$ can be interpreted as local scaling dimensions,
this implies the dense regions are effectively two-dimensional structures.
This will be further discussed in Section~\ref{obscomp}, in comparison
with reported multifractal indices for observational and simulated data.

Furthermore, based on the relative consistency of the dimensions, one could
conclude that these
are signatures of the {\it construction algorithm}, and independent of
the space in which they are built.  Note that the fractal dimension of 
two-dimensional
Apollonian Packing is reported to be in the range $D_F \sim 1.31$ \cite{apppack},
thus perhaps the $D_0 \sim 2.6-2.7$ is a signature of the 3-dimensional packing.
This figure should be compared with that of Reference~\cite{app3d}, which 
reports a value of $D_F \sim 2.4$ for 3-D Apollonian Packing (although 
the author of \cite{stern} notes this to be ``crude'').
This notion of construction mechanism identification will be further discussed
later in this paper.

Based on conclusions drawn out in Reference~\cite{mythesis},
the dimensions in Tables~\ref{pos1} and \ref{neg1} are obtained via
box counting in the flat projective subspace.  This projection
is justified, however, since the overall angular
extent is $2\omega_{\rm max} = \pi/5$ (see Section~\ref{geodesics}),
which constitutes a rather small
portion of the hyper/pseudospheres.  
It has been shown in \cite{mythesis} that 
curvature effects become minimal at about this angular
extent.  So, in some sense, one cannot expect to see {\it any} signs of
curvature for such a limited library.

It should be noted that since the libraries represent only the first-level 
in the recursion, the depth ranges only to less than 3 orders of magnitude, 
beyond which point
the scale approaches the size of the smallest spheres and/or the mean
interparticle distance.  Adding a second level of recursion could help to 
better define the overall dimension, although this can lead to significant
increases in the computation time required to perform the operation.
Future faster processors combined with efficiently-written algorithms will
certainly be able to handle such a task.  This work is currently underway
by C.\ C.\ D.\ {\it et al.}

From a formulaic approach, it is interesting to note that homogeneity
is preserved on the first scales of iteration in the box counting 
procedure.  That is, cubes of side $\sim d, d/2, d/4$ 
(where $d \sim \pi/10$ is the radius of
the set) completely cover the distribution, giving counts of 8, 64, and
(approximately) 256 respectively.  
This ``space--filling'' quality of the packings is
consistent with the design of
the Swiss Cheese cosmology ({\it i.e.} intended to satisfy the 
Cosmological Principle on the largest of scales).

As a quick check, these statistics are in relative agreement 
with fractal dimension
calculations via alternative codes or methods.  A shareware piece
of software known as {\bf fd3} was obtained \cite{fd3}, which calculates the 
box dimension of an N-dimensional array of an arbitrary number of 
coordinates.  The corresponding box-counting algorithm is based on the 
method of reference~\cite{fast}.  
The {\bf fd3} software rather consistently pegs the dimension of the packings
at $D_0 \sim 2.5-2.7$, with mild variation in the correlation dimension
$D_2 \sim 2.4-2.6$, in rather good agreement with the above results.

\section{Comparison to Observational and Simulated Data}
\label{obscomp}
The results of the previous section suggest that the reported $D_F \sim 2$ of 
large-scale clustering (Table~\ref{galaxydims}) cannot be reproduced by the 
PSC models with any level of certainty.  This could imply either that the PSC
model under consideration is limited in its predictive power, or equally that
the cited $D_F \sim 2$ is incorrect.
It may be that sample size plays a critical role here: the actual surveys have 
considerably smaller counts than the PSC models considered herein.  Note
that smaller models showed lower fractal dimensions, so while this could
indicate different scaling estimates at lower ranges, it does not support
the extension to larger scales.  It is not unreasonable, then, to expect
different scaling behavior over different ranges.  

Various multifractal analyses of the available redshift catalogs have
been performed, but unfortunately, there is a relative lack of these
clustering analyses in the literature, as compared to the more traditional
correlation function (single fractal dimension) investigations.  
However, the majority of 
those which can be found tend
to indicate commonalities in the data (with the exception of studies
such as those discussed previously in \cite{eso}, although this was
not a full multifractal analysis).  
Recall that values of $D$ obtained from the standard correlation analyses
should generally correspond to $q=2$ (not $q=0$), 
although in some cases this tends
to be a point of confusion among some authors and the 
related interpretations of analysis data.  Analyses which are purely
monofractal (correlation or conditional density) quote only a single exponent
with no regard for the $q$-index, which makes for sometimes ambiguous 
cross-method comparisons.  It was first noted in
\cite{jones1} that from a multifractal analysis of the CfA cluster,
there existed a mismatch between the box dimension ($D_0 \sim 2.1$)
and the widely reported correlation dimension ($D_2 \sim 1.3$), indicating
that the large-scale clustering was inadequately described by a single
scaling dimension.  

Further investigations of CfA1 were reported in \cite{apj332,apj357} in 
which it was determined that the spectrum of generalized
dimensions ranged from $D_0 \sim 2$ to $D_{\infty} \sim 0.6-0.7$. 
More recently, multifractal dimensions were extracted for
the CfA2 survey \cite{aa344,aa370},
as well as the Las Campa\~{n}as survey  \cite{aa370} (see references
therein for details on each catalog).  Again, a similar
trend in dimensions was observed, although instead citing 
correlation dimensions
with values $D_2 \sim 2$ for $r \sim 5-30~$Mpc, 
and strong local clustering behavior
of $D_{\infty} \sim 1$ (noting that the discrepancy in $D_2$ values
from those reported previously may be due to the spatial extent of
the catalog considered \cite{aa370}). 
It was further noted in \cite{aa370} that the multifractal
behavior was only observed over the aforementioned distance scales, 
with $D \sim 3$ beyond ({\it i.e.} a transition to homogeneity ensues,
again contrary to ``no transition'' conclusions of \cite{piet1}).

Similarly, cold dark matter (CDM) N-body simulations of gravitational 
collapse have also been
studied, with essentially similar conclusions being drawn.  
These are of particular interest due to their natural connection to
hierarchical clustering growth from small initial mass/density perturbations
in the early Universe.  The main
difference, however, is that while observational data tend to 
show ``fluid'' multifractal
structure, the latter simulations seem to display 
moreover a ``bi-fractal'' scaling \cite{apj394,borgani,dubrelle,colombi}.
Reference~\cite{apj394} presents a comprehensive set of dimension estimates for
such simulations at various epochs ({\it i.e.} varying cosmological 
scale factor), as well for increasing biasing levels (the biasing level
implies that clusters will only form for peak densities above a certain
threshold value, generally a multiple of the standard deviation of the
mass fluctuation \cite{padman}).
The authors demonstrate that any number of box dimensions between $1.2 - 2.7$
are allowable, depending on the epoch and biasing levels ({\it i.e.}
sensitivity to higher density fluctuations).  Higher dimensions
are observed at earlier epochs (when the matter distribution is still
predominantly homogeneous), with the extreme lower dimensions occurring
for high biasing.  For all simulations 
considered, however, a consistent signature
of the evolution is $D_{q} \rightarrow 1$ for larger $q$, due to the
gravitational in the overdense clustering regions ({\it i.e.} non-linear
density perturbation regime; see {\it e.g.} \cite{peebles2,padman} for a 
review).

References \cite{dubrelle,colombi} support this claim for
behavior in various CDM N-body simulations, where the high dimension
$D \sim 2$ corresponds to larger-scale correlations, with 
$D \sim 1$ behavior at smaller scales (and no interim scaling).  This
motivates debate as to how well even the N-body simulations reproduce
observation.
In reference~\cite{borgani}, which studies the reliability of various
fractal dimension estimation methods, the authors discuss several random
process simulations which reproduce both the multifractal spectrum of
observation, as well as the bi-fractal nature of such N-body simulations.

Thus, in addition to the mismatch between low $q > 0$ dimensions, 
a further striking difference between the PSC models and the above
cited results from
both observational and simulated data are in the $D_q$ values
for $q \rightarrow \infty$.  This is attributed again to strong 
gravitational clustering effects at small scales which 
dominate the densest portions of the structures \cite{borgani,apj394},
whereas the dimensions obtained in the packing library analyses herein 
tend to indicate 
that $D_q \sim 2$ for large $q$.  The discrepancy is undoubtedly due again
to the formation mechanism of the structure in question.  The N-body
simulations (and similarly the real local clustering effects) 
are evolutionary and dynamic, and hence their overall structure 
depends on the integrated gravitational interactions between the individual
particles coupled with the scale expansion of the Universe. 
Thus, the densest structures will be ``stringy'', with one
preferred direction being singled out.  This is consistent with the
observation of galactic ``void-filament'' structures, {\it i.e.} 
linear aggregate structures which cluster on surfaces. 

The PSC algorithm, in contrast, creates the structure ``on-the-fly'', with no 
consideration of gravitational interactions.  Again, its construction paradigm
is purely geometric, based on volume-optimization, 
and the strongest clustering will occur
by smaller spheres on the surfaces of the largest spheres.
So,  $D_{\infty} \sim 2$ can be interpreted as
the densest regions clustering on {\it surfaces}, or having clustering
behavior that has only two preferred directions.  This is certainly a
reasonable statement, since by the very nature of the packing routine,
one should expect the smallest spheres to cluster
on or near the surfaces of the largest inscribed spheres (one can obtain
a sense of this from Figure~\ref{packgif}).   
These dense regions, on the other hand, are
reminiscent of the Great Wall and similar structures
observed in such redshift surveys as CfA (see {\it e.g.} \cite{geller}).

As a quick clarification, the use of the term ``preferred'' in this case is 
used in an opposite fashion to the usual discussions of homogeneity.  
Note that $D_F = 3$ would be described as structure having three 
``preferred'' scaling directions ({\it i.e.}
instead of no preferred direction -- simply put,
all directions are equally preferential).

The obvious mismatch in $D_q$ spectra between PSC and observation/simulation
is discouraging, but it is not to say that the PSC model is inherently flawed.
Rather, these figures
should be seen as accepted limitations to the procedure, considering the
accuracy of the data and associated statistics.
The PSC models implicitly preserve the Cosmological
Principle and Weyl flatness, while N-body simulations actively break these
symmetries (or at the very least, do not concern themselves
with consistent maintenance of this condition throughout the evolutionary
process).   

It is perfectly reasonable, however,
to suggest that the two differing models are {\it complementary}, rather
than in opposition.  PSC shows large-scale homogeneity, preserves the
CP, and displays strong clustering in two dimensions surrounding 
voids.  All that is really missing is the filamentary structure.
That is, the N-body simulations could always take
place {\it within} any particular sphere of the packing. 
Recall that although the Weyl condition
is satisfied on the large scale, in accordance with the FRW dust solutions, 
but locally (within the sphere) it can be violated ({\it i.e.} the vacuum
is Ricci flat, but a non-vanishing Weyl is required to maintain overall
curvature).  Thus, the Weyl constraint provides a necessary boundary 
condition for the preservation of the CP.  A hybrid of different clustering
paradigms could help to yield a more realistic approximation to the observed
structures and theoretical constructs.
Equivalently, if the N-body simulations alone are truly bi-fractal, then
combined with a PSC-like structure, one can obtain a more realistic
model with which to compare observational data.

The fact that this discrepancy between dynamical and ``instantaneous''
formation paradigms can be readily signed through the multifractal
analysis, and that the Universe more closely resembles the N-body
data, can lend a sense of relief to philosophers.  That is, this is a
sure signal that the Universe has been evolving, or at the very least,
was not created on-the-fly in the manner of the PSC!  Thus, again, it
would be interesting to study the multifractal scaling behavior of an
N-body simulation nested within a PSC library, left to evolve over
time.  If {\it this} were to match observational data, the result
could have profound impact for early Universe dynamics (not to mention
the philosophical implications of a pre-determined structure in the
beginning).

Along the same lines, perhaps the observational $D \sim 2$ of the 
galaxy clusters is actually 
a measure of this quantity, constrained by the relatively small
sample space of candidate galaxies with respect to the estimated size
of the Universe.  Hence, the PSC could conceivably
represent a ``large-large'' scale structure of the Universe, that is,
much larger than the current observational limits.  This would then
be consistent with the above characteristics of observation and theory.

A brief mention is in order for
another earlier hierarchical cluster model of related interest, that of 
Soneira and Peebles \cite{peebles3}.  Within a volume of space, a sphere 
or radius $R$ is inscribed, and within that sphere are placed $n$ spheres
of radius $R/\rho$.  In turn, within each of those spheres are placed
$n$ new spheres of radius $R/\rho^2$, and so forth.  It is interesting
to note the conceptual similarities between this model and the SC packings,
both of which are based on instantaneous positioning of spheres in a volume,
and not on any time-evolution paradigm.  The Soneira-Peebles model
is not explicitly multifractal, but rather a simple monofractal of dimension
$\log(n)/\log(\rho)$.  The corresponding multifractal spectrum is effectively
flat, such that $D_q = D_F \; \forall q$ for $r \in (R/\rho^{n-1},R)$,
where $n$ is the recursion level \cite{apj357}.  This, however, bears
little resemblance to the PSC spectra, nor for that matter to 
observational data.

\subsection{Comparison with Random/Ordered Distributions}
\label{random}
Since the results do not suggest any method of discrimination between the
three classes of models, it is important to address the question of 
how the multifractal structure of the packings compare to those of other
random distributions of points.  As a test, several toy random spherical 
distributions were populated using the C library function {\bf drand48()},
which on each call generates a pseudorandom floating point number 
in the range $[0,1)$.  Since curvature plays minimal or no role in the
resulting spectra, for conciseness comparisons will be made only to
the flat libraries.

One set of distributions is designed to be completely randomized, such
that each $(x,y,z)$ coordinate is generated randomly.  A second, dubbed
``linear-random'', randomly generates $(r,\theta,\phi)$ coordinates in
the usual range.  A third represents a completely uniform spacing of
points which fills the appropriate volume.  These three test sets, along
with a random packing library, are displayed in Figures~\ref{randsets1}
and ~\ref{randsets2}.
Since no apparent distinctions can be made between the flat, 
open, or closed universe distributions, only the flat library SCF3 will be 
addressed in this section.

The simple Box dimension ($D_0$) shows mild differences between each set,
and in certain cases fails to accurately distinguish between them.  However,
the larger $q$ values do indeed show large variations in the overall
structures.  In particular, note that the linear-random distribution gives
a box dimension of $D_q \sim 2.8$, but rapidly decreases for $q > 0$ to
a value of $D_q \rightarrow 1$ for $q \rightarrow \infty$.  This exemplifies the
ability of the multifractal spectrum to pick out anomalous structural
qualities.  Although the coordinates $(r, \theta, \phi)$ are populated
randomly, the {\it volume element} of the space is not.  
For randomly selected $(x,y,z)$, the volume element $dV = dx\:dy\:dz$ is
itself purely random, while randomly selected $(r,\theta,\phi)$
coordinates will yield a radially-weighted volume element $dV = r^2
\sin\theta\:dr\:d\theta\:d\phi$.  The overall clumping will reflect
this $r$ preference, and hence will appear ``pseudolinear''.
Since each set fills the volume almost homogeneously,
and as such ``fools'' the box counting algorithm into a misrepresentation
of the entire structure.  This LR set can be compared to 
the local behavior of the N-body simulation dimensions discussed in 
Section~\ref{obscomp}, both of which are dominated by purely radial ({\it e.g.}
gravitational) clustering behaviors at small scales.  In fact, it is
interesting/amusing to note that for this set that the correlation dimension
$D_2 \sim 1.9$, further suggesting that
this test data could be considered a potential candidate for large-scale 
galaxy clustering (although purely in jest; this behavior appears more 
bi-fractal, due to the quick decrease to $D_q \sim 1$, and most likely
lacks additional structural features beyond the $D_q$ spectra).

The $D_{\infty}$ value signals the linearity of the
densest clustered regions.  The randomly populated $(x,y,z)$ space (RD) is
more reflective of a truly random sample, as indicated by the dimensionality.
Note that ideally, the set should have $D_q = D_0 = 3\; \forall q$, but
the variance can be understood to be software and data set limited (the
same would be true of UD, which is just a variant of the RD). 

\subsection{Ribeiro's Tolman Swiss Cheese Cosmology}
\label{ribeirowork}

As previously mentioned, the structure of a Swiss Cheese-like cosmology
is a natural choice if one adheres to the existence of local inhomogeneity
while adhering to the Cosmological Principle.  Similar solutions have in
fact appeared in the literature before, perhaps one of the most thorough 
being the model presented in the series of
papers by Ribeiro \cite{ribeiro1,ribeiro2,ribeiro3}.  It is interesting
to compare these results -- which discuss a
relativistic fractal cosmology -- to those of the current paper.  
Ribeiro's model is very similar in structure to
the present Swiss Cheese model considered herein, matching Tolman solutions with
FRW dust solutions and integrating local density distributions along the
past light cone to calculate the observed fractal dimensions.  For a variety
of classes of solutions, the author finds fractal dimensions which range between$D_F = 1.3-1.7$ depending on the model type under consideration.
(subject to the constraint of obeying the de Vaucouleurs' density
power law).  The interested reader is directed to the aforementioned
citations for further reading.

\section{Curvature Considerations}
\label{kconsiderations}
\subsection{Evaluation Along Geodesics in $k = \pm 1$ Spaces}
\label{geodesics}
Ideally, one must be careful when evaluating radial distances in non-flat
geometries.  Although it has been argued that the curved geometries considered
herein possess ``essentially-flat'' characteristics due to their small
angular extent, it is worth a quick check to see if there might exist any
overt differences between the flat box estimates and those which do consider
curvature effects.

The spatial portion of the FRW metric (\ref{frwmetric}) can be written as 
\bea
d\sigma^2&=&d\omega^2 + S_k(\omega)^2 \left[ d\theta^2 + \sin^2\theta\;d\phi^2
\right]~,
\label{frwmetric2}
\eea
where the Cartesian embedding coordinates for the corresponding hypersurfaces
are \cite{mtw}
\bea
x&=&S_k(\omega)\sin(\theta)\cos(\phi)  \nonumber \\
y&=&S_k(\omega)\sin(\theta)\sin(\phi)  \nonumber \\
z&=&S_k(\omega)\cos(\theta) \nonumber \\
w&=&T_k(\omega) \nonumber \\
\label{fourcoords}
\eea
Here, $T_k(\omega) = \cos(\omega)$ for $k=+1$ ($\cosh(\omega)$ for $k=-1$),
and the radius of curvature has been set as $R=1$.

Realistically, one must compute the geodesic distance between
points in the embedded space, instead of the straightforward Euclidean
distance.  For the positive and negative spaces considered herein,
Equation~\ref{densrecon} is evaluated along
angular geodesic distances $r=\delta$, which is extracted from the
usual inner product in the space between vectors
$(\omega_1,\theta_1,\phi_1)$ and $(\omega_2,\theta_2,\phi_2)$,
\bea
\cos \delta&=&\cos(\omega_1)\cos(\omega_2) + \sin(\omega_1)\sin(\omega_2) [
\sin(\theta_1) \sin(\theta_2) \cos(\phi_1-\phi_2)  \nonumber \\
& & + \cos(\theta_1)\cos(\theta_2) ]~, 
\eea
for $k = +1$.  The analogous expression for $k = -1$ obtained by the
usual replacement for $S_k(\omega)$ and the appropriate sign change, 
\bea
\cosh \delta&=&\cosh(\omega_1)\cosh(\omega_2) - \sinh(\omega_1)\sinh(\omega_2) 
[\sin(\theta_1) \sin(\theta_2) \cos(\phi_1-\phi_2) \nonumber \\
& & + \cos(\theta_1)\cos(\theta_2) ]~,
\eea 

The scaling dimensions may then be evaluated with ``bottom-up'' estimation
techniques along the geodesic.  The  
conditional average density $\Gamma(r)$ of points in the set, defined
as
\beq
\Gamma(r) = \frac{1}{N}\sum_{i=1}^{N}\frac{1}{A^s_i(r)} \frac{dN_i(r)}{dr} \propto r^{D-3}~,
\label{conddens}
\eeq
estimates the average change in number of points within a spherical shell of
radius $r=\delta$ (area $A^s(r)$).  The shell area 
$A^s(r)$ will depend on the value of
$k$, however these effects are exceedingly small and can be ignored.
This is reported to be a better estimation tool than the standard correlation
function $\xi(r) = (r/r_0)^{D-3}-1$ (where $D_F = D_2$), 
which makes {\it a priori} assumptions
of homogeneity with regard to the set under consideration (for an
inhomogeneous fractal object can lead to spurious results) \cite{piet1}.
Of course, disagreements between dimension estimates from these two methods
can also arise if the distribution is not a pure monofractal.
If the set in question is fractal,
then the number $N(r) \propto r^D$, and thus $\Gamma(r) \propto r^{D-3}$,
which provides a good measure of the fractal dimension $D_0$ 
(and is frequently used in the cited references).  Evaluating (\ref{conddens})
over $N\sim 2000$ points per set or less, with an equivalent $R_{\rm eff}
< 0.10$, one can obtain figures in good agreement with
the cite box counting estimates.  
%For example, the SCF sets (Euclidean $r$) 
%give dimension estimates between 2.5-2.6 ($\pm 0.1$), with SCP 
%(geodesic $r$)
%between 2.6-2.8 ($\pm 0.1-0.2$), and SCN between 2.6-2.9 ($\pm 0.1-0.3$).  

The plots of Figures~\ref{cdensfig1}-\ref{cdensfig3}
show $\Gamma(r)$ for sample libraries of each
curvature, as compared with a base slope of $2.8$.  
The fit itself is somewhat dependent on the
choice of points, but all seem to suggest a similar slope.
The smaller population sets show higher variability in the convergence
of points to linearity, but this effect is smoothed out as the set
grows in size.  Note that the
flat libraries tend to yield slightly smoother trends than the positive
or negative sets.  While this could
be a curvature effect, it is unwise to make such an assertion without further
investigation.  The apparent non-linearity of each set below $r \sim 0.01$
is a recognized artifact of finiteness in volume-limited samples such
as these, as the radial distances drop
below the range of statistically-significant clustering \cite{piet2}.
Hence, the cited dimensions are obtained for the linearity which ensues
to the right of this ``peak''
This method is useful as a self-check here, since the 
calculated dimensions tend to support those obtained via the box counting
method.  Since it cannot be used to measure additional multifractal structure, 
however, its uses have been exhausted.   

\section{$D_q$ for $q < 0$}
\label{negqs}
The value of the parameter $q$ need not be restricted to positive integers,
nor in fact need it be restricted to integers.  The $D_q$ are well-defined
for all $q \in \cal{R}$.
Whereas the $D_q$ values for $q > 0$ represent the scaling behavior
of increasingly dense regions, those for $q < 0$ correspond to the
scaling of {\it under-dense} regions.  
Both the analyses of observational data and N-body simulations mentioned
in Section~\ref{obscomp} suggest that $D_q \rightarrow 3$ for 
$q\rightarrow \infty$, indicating that in all cases the least populous
or dense regions scale effectively as homogeneous distributions.

Numerically, obtaining $D_q$ values of $q < 0$ 
can be a very difficult quantity to estimate via box counting, since 
calculations becomes severely dependent on the finite size and 
population of the data set \cite{borgani}.  Such behavior
is observed herein -- the associated error estimates grow significantly
for $q \geq 0$, particularly for 
values of $q < -1.5$ or so.  For example, the box counting algorithm
yields $D_q \geq 2.7~(0.3)$ for $q = -2$ for SCF3.  Similarly, SCP5 
shows $D_q \sim 2.8~(0.2)$, and $2.7~(0.3)$ for SCN3.  The other libraries
considered demonstrate roughly similar behavior, although below these
values of $q$, the fit errors quickly grow.   
In fact, in certain cases the estimated $D_q$ values begin
to {\it decrease} for sufficiently small $q$, but with significantly increasing
error (for all points fit).  This could indicate that accurate determination 
of such generalized dimensions is extremely dependent on the choice of points
returned from the calculation.

There are several other means by which one can more accurately compute these
values.   A common method used in astrophysical analyses is 
the {\it density reconstruction} algorithm (see {\it e.g.} \cite{apj357,apj394}
and related references), which determines the minimal 
radius $r(p)$ around a point for which the probability is $p$ of finding 
$p\:N_{\rm tot}$ points.  The corresponding partition function
\beq
W(\tau,p) = \frac{1}{N_{\rm tot}} \sum_{i=1}^{N_{\rm tot}} r_i(p)^{-\tau} \propto p^{1-q}~,
\label{densrecon}
\eeq
Recall that the multifractal spectrum $\{D_q\}$ is defined by the relation
$\tau \equiv \tau(q) = (q-1)\:D_q$, thus the
corresponding generalized dimensions may be obtained accordingly.
This method is claimed to converge well for $q \leq -1$, but is still
applicable for a small range of positive $q$.

The calculation is somewhat computationally intensive, since ideally
the radial calculations must be made for {\it every} point in the set.
Generally, a smaller (random) sampling of the set can be used to evaluate
the partition function.  Sample calculations via implementation of this 
algorithm have been implemented for a range of probability values
$p \in [0.01,0.1]$ (similar to the range used in \cite{apj357}), 
for points whose inscribed radius is not overly
large (else the calculation becomes saturated and useless), 
which generally includes up to several thousand points.  In the cases
of $k = \pm 1$, these are evaluated out along the geodesics (although
as has been established, this complication likely isn't necessary).

Figures~\ref{dqlibs1}-\ref{dqlibs3} show sample $D_q$ plots for various libraries considered
herein.  The curves are
shown to demonstrate the general trend in $D_q$ values, since there is
sometimes a mild mismatch between the two methods (although within the
box counting error).  Thus, explicit points and error bars have been 
suppressed for the time being.
The errors are roughly 0.1, and so the curve can be assumed to be within
1- or 2-$\sigma$ of the actual values.
Initial results indicate that the 
scaling dimensions of the packings for small $q << 0$ approach $D_q \sim 3.0$,
as was suggested by the box counting method.  The rate at which the values for
individual sets in each class (flat, positive, negative) 
approach the limiting value
show some mild variation, but like the cases for positive $q$, this behavior
could simply be due to the size of the data set under consideration ({\it e.g.}
SCF1 shows markedly lower values than the rest, perhaps due to its smaller
population size).  

The density reconstruction method is also applicable to (small) values of
$q > 0$, and it is noted that this provides good agreement with the
box dimension estimates of Section~\ref{swisscomp}.  In particular, the
estimates for $q \sim 2$ tend to be closer in agreement to the 
fits of Tables~\ref{flat1}-\ref{neg1} to within the cited error, 
thus providing yet another self-check for the estimates therein.  

In certain cases, the $D_q$ value does not rise much above that of
$D_0$, however it is uncertain whether or not this is an actual artifact of
the data set, or rather calculation anomalies.  This does not appear to
be an explicit artifact of curvature, since it occurs for packings belonging
to each class.  It could be that these values of $D_q$ can serve as some
variety of identification for varying initial conditions or the general
distribution of sphere sizes (see \cite{cheese1}).  That is,
while $q > 0$ is a measure of the strongest clustering, which can be interpreted
as an overabundance of local mass at small scales, $q < 0$ is to a certain
extent a measure of open space, which certainly could vary from library
to library.  It could also be some manifestation of curvature on the
construction paradigm at the largest levels \cite{mythesis}.
Certainly, further investigation into the behavior is ideally warranted.

Coupled with the box counting results from before, these figures seem
to consistently suggest an asymptotic value approaching
$D_{q \rightarrow -\infty} \sim 3$.  Such behavior has the simple 
interpretation that the regions of least clumping scale homogeneously, and
thus no explicit structural information can be extracted from these values.  
Similar behavior is reportedly observed for the N-body simulations
discussed previously, as well as the analysis of observational data from
the diverse galaxy catalogs.   Thus, differentiation of formation
models seems to become
even more ambiguous for this range of scaling.  The strong clustering behavior
for $q \gg 0$ gives a much more intuitive and exploratory glimpse of the 
inherent structure qualities of the set.  

\section{Biasing Counts by Luminosity / Mass}
\label{wtmass}
As previously noted, the estimated scaling dimensions for the packing
libraries are high due to the density of points per allocated volume.
Another issue is the unbiased nature of the counts.  Simply put, the
libraries are geometric configurations which satisfy the relativistic
field equations.  Each point is assigned a mass, which is a function
of the inscribed radius within the constant density field, dependent
on the curvature of the spatial manifold.  However, to date this value
has not been factored into the counting scheme.

The observed three-dimensional distributions are interpolated from 
measured catalogs -- what one sees is what one gets.  However, it is
not unreasonable to suggest that what one sees is {\it not} what is
actually there!  The notion of ``missing mass'' is a 
recurring theme in many cosmological endeavors, ranging from 
dark matter issues, to low luminosity objects.  
Furthermore, it is common-place in 
observational astronomy to simply excluded objects whose luminosity
is below a given threshold, {\it irrespective} of whether or not
it is visible.  If it does not meet the selection criteria for the 
catalog in question, it is omitted altogether.  Furthermore,
the issue of luminosity biasing or segregation effects is one which is
frequently raised in the debate surrounding the $D \sim 2$ over all
length scales.  For example, it has been 
suggested that luminosity segregation favors brighter galaxies at greater
distances, which consequently implies a biasing toward stronger clustering,
which could skew the actual measured dimension from homogeneity to
some different scale (see \cite{apj394} and associated references therein).

The mass-to-light ratio for galaxies is somewhat dependent on factors
such as age, morphological type, and so forth.  It seems
reasonable to impose a mass cutoff in the packing libraries as a first-pass
gauge of this effect.   That is, can one impose constraints on the range
of lower mass cutoffs from which one can extract a dimension of $D\sim 2$
for the packing libraries?  For most galaxies of
a particular class, this ratio is roughly a constant, {\it i.e.}
$M(L) \propto L^\beta,~\beta \sim 1$.  For simplicity,
such a relation will be assumed for the analysis herein, and thus the
results can provide a good ``boundary'' for what one might expect in 
realistic observational circumstances.
Since the initial matter field is assumed to be of constant density
$\rho_0$, the mass $M_i$ of the $i^{\rm th}$ sphere may be calculated 
by the usual relations
\beq
M_i = \frac{4 \pi \rho_0}{3} R_i^3~~,
\label{fmass}
\eeq
\beq
M_i = \pi\rho_0\left(2R_i - \sin(2R_i)\right)~,
\label{pmass}
\eeq
\beq
M_i = \pi\rho_0\left(\sinh(2R_i) - 2R_i\right)~,
\label{nmass}
\eeq
where $R_i$ is the inscribed (angular) radius of the sphere, for 
flat (\ref{fmass}), positive (\ref{pmass}), and negative curvatures
(\ref{nmass}).  The latter expressions are obtained by integrating
the FRW volume element for the respective values of $k$.  Since $\rho_0$
is arbitrary, one can set $\rho_0 = 1$ WLOG.

Figure~\ref{dqnegs} shows the mass-reduced $D_q$ spectra for  various
reduced libraries, using a mass cutoff approximately $0.01\%$ 
that of the largest sphere mass.  In terms of the assumed mass-to-light 
ratio, this logically amounts to a similar scaling for the object's luminosity. 
The reduced data sets contain roughly 10\%
of the original number, but still accounts for over 90\% of
the total enclosed mass.  Similar results are obtained for the other packing
libraries using a similar cutoff.  This indicates that the majority of the
spheres are relatively small in mass, and thus by the mass-to-light ratio,
would most likely be much fainter relative to the remaining points. 
Hence, the smaller masses serve to ``smooth out'' the overall matter 
distribution, leading to the saturation of homogeneity observed in the
upper-most levels of the box counts.  

Note the somewhat wide variation in limiting values of $D_q$ for
negative $q$.  Although the data confirms sparser clustering in these
regimes, the asymptotic value which $D_q$ assumes tends to be somewhat
dependent on the library.  Initial results seem to suggest that
there is no preferential pattern for specific geometries, with both
possessing slow and fast approaches to a higher limiting dimension.
It may be the case, as discussed previously, that
further investigation into the $q < 0$ regime could potentially
shed light on the associated geometry, although a better understanding
of the sensitivity of the method to population sizes is required.  
The authors of \cite{apj394}
similarly suggest that $q <0$ values could help discriminate between
differing initial power spectrum perturbation conditions for 
N-body simulations.

Table~\ref{allbias} displays the results for the associated cutoffs for
each the libraries of Tables~\ref{flat1}-\ref{neg1}.  Again, note that
there is no explicit signature variation in the 
estimates between curvature cases.
The SCP libraries show slightly higher $D_0$ and lower
$D_{\infty}$ values than the SCF and SCN packings, albeit all equal
to within the associated error.
The same mass cutoff has been applied to each case, even though there is
mild variation in the overall masses for each library (depending on the
packing number).  However, this is most likely a small consideration which
does not significantly affect the end result.  This variation could also
be due to volume limitation constraints which exist in the closed manifold,
but not in the flat and open cases.

%In terms of the Weyl Tensor constraint discussed in Section~\ref{grcp},
%the results of Table~\ref{allbias} have important consequences for the
%PSC model.  The large number of small masses serve in some sense as filler, 
%but with the very important purpose of providing the grounds for which 
%the vanishing Weyl Tensor is preserved throughout the space in question.

Thus, it would appear that 
the use of luminosity cutoff can potentially bring the box and correlation
dimensions of the libraries in closer agreement to the reported values of
reference~\cite{piet1} (albeit with larger fit error), 
which suggests that such a biasing mechanism could
help to rationalize the aforementioned discrepancy between the PSC models
and observation (if correct).  Such luminosity biasing in Abell clusters
was discussed in detail by Bahcall and Soneira, who showed
a mismatch in the galaxy and cluster spatial correlation lengths of at least a 
factor of 5 \cite{bahson} as well as a strong dependence of the correlation
function on cluster richness.  The more recent analysis of the SSRS2 redshift 
survey presented in \cite{benoist} further supports the conclusion 
that there exists a strong connection
between statistical clustering and luminosity, in particular weighted toward
bright galaxies ({\it e.g.} $M \leq -21$) over fainter ones.  The authors
comment that these results are largely inconsistent with current theoretical
models, however this paper indicates that the PSC could offer a resolution
to this observed effect over other clustering paradigms.

Furthermore, this can help support the notion that 
observed galaxies (luminous matter) can form a fractal 
distribution (subset) within a largely
homogeneous matter distribution, much the same way the 
distribution of mountain peaks is fractal, while the Earth itself is
largely spherical \cite{frachom}. 
Observationally, similar cutoff biasing has been
recorded, and that the associated fractal dimensions tend to rise for
a decreasing galaxy luminosity threshold \cite{piet1,frachom}.  Thus, 
the ideas presented herein are consistent with reported 
astrophysical procedures,
and could always be used to help provide insight into 
similar ``missing mass'' investigations.
As with the fully unbiased model, however, the values
of $D_{\infty}$ for all libraries considered tend to cluster around
$D_{\infty} = 2$, which again does not match observational and N-body simulated
models (whose $D_{\infty} \sim 1$, as previously discussed).  

However, some caution must be exercised in interpreting such results.  
This trend may be moreover a statistical manifestation influenced by
a different linear fit choice than an actual structural change in the set.
In fact, by neglecting the smallest box size count in the fits for
the higher cutoff biased sets, the $D_0$ 
values can be shown to rise to $\sim 2.4-2.5$ with smaller fit
error, suggesting that the dimension is potentially higher than indicated in 
Table~\ref{allbias} (although still lower than the estimated $D_0 \sim2.7$ 
for the whole library).  
There is also a marginally larger range  between the
differing dimension estimate techniques for the mass-biased libraries.
For example, the density reconstruction method
tend to yield values of $D_q \sim 2.3-2.5$ or so for $q \sim 0$
(see the trends of Figure~\ref{dqnegs}),
indicating potential limitations of the box counting technique.

Figure~\ref{mcvar} demonstrates the trend in dependence of the $D_q$ values
on cutoff size, both for fixed number of iterations for all cutoff levels,
as well as adjusted numbers for smaller set populations.  The estimates 
remain largely unaffected until a cutoff
of $> 0.001\%$ is reached (leaving about 20\% of the spheres and 96\% of
the total contained mass), after which there is a relatively quick drop
in $D_0$ with respect to $D_{\infty}$ for the fixed number of iterations.  
Whereas, adjusting the number of box levels (reducing by one) helps to
correct the curve to a higher estimate.
Beyond about $0.04\%$, the dimensions
approach roughly the same value $D < 2.0$, suggesting a type of monofractal
behavior.  At this point the population of the set has been
reduced to 2\% of the original, thus the statistics could start to 
become skewed by finiteness effects due to a larger average point separation.

The box counting technique has at times been criticized in the literature
for being too sensitive to discreteness effects, especially when the
population size is small (causing spurious results or underestimates
of the actual scaling dimensions) \cite{dubrelle,colombi}.  
In other cases, box counting
is hailed as a quite robust and stable method for estimation of the
associated statistics, and furthermore the density reconstruction
method is claimed to give dimension overestimates for limited size
sets \cite{borgani}.  Thus, there is a certain level of disagreement and
confusion in the community with respect to the utility of any of these methods.

So, when few data points are available such as in this situation, 
it is probably best to take some variety of
average dimension as calculated by differing methods.
Although the measured $D_0$ dimensions do drop when lower masses are
discounted, the actual magnitude of the drop seems sensitive to the 
size of the data set (and measurement method).   Whether
or not this represents a skew in the measured dimension as a result
of luminosity biasing is somewhat unclear.
Thus, such biasing results are more susceptible to statistical 
anomalies (highly dependent on the choice of points in the box count fit) 
than are the full models, and care must be taken in their use.  
It is, however, worth noting that the $D_{\infty}$ values remain relatively 
consistent at $D_{\infty} \sim 2$, indicating a certain robustness of the 
data.  This is a {\it definite} signature of the packing structure.

From a physical point of view, one must recall that the PSC
models are merely weighted points.  As mentioned in a previous section,
one could imagine replacing each point with another type of set, such
as an evolutionary N-body simulation, which could then conceivably yield
finer clustering effects (and potentially yield a $D_{\infty} \sim 1$
behavior).  If the dimensions of Table~\ref{allbias} are
actually statistical properties of the luminosity-biased sets, then
it could constitute a possible solution to reconciling the mismatch
with observational data.

\section{Future Considerations}

The basic premise of the packing algorithm is to ensure preservation of
the Cosmological Principle via maintenance of Weyl flatness.  Although
this is done by compressing the matter contained within a spherical region,
there is {\it a priori} no reason for this choice (apart from the motivation
of gravitational collapse).  Suppose instead the matter were {\it expanded}
to lie along the spherical shell described by the inscription.  This
would create a thin sheet of matter, consistent with the two-dimensional
structures complemented by voids, evident both from observation as well
as the multifractal analysis.  The matter could then be allowed to 
coalesce by some gravitational mechanism on the sheets, yielding the
appropriate linear structures.  Alternatively, the packing algorithm
could randomly choose between collapse and expansion for the spheres,
and in the case of a collapse, the interior matter is allowed to 
cluster via N-body behavior.
Whether or not such models are physically realizable is unknown, but 
they could nevertheless provide yet another model from which to study
and reconcile the observed large-scale structures in the Universe,
and as mentioned, stand to uphold the Cosmological Principle on many
levels.  

Furthermore, the use of the monofractal dimension can frequently short-change
the characterization of a clustering set, since it is possible to have 
vastly differing structures which possess the same base dimension $D_F=D_0$.
Although the multifractal spectrum can help differentiate such situations,
a different consideration is the {\it lacunarity} of the set.  This measure,
often associated with the texture of a fractal, provides an
estimate of the ``voidness'' (rather than ``clumpiness'') (see {\it e.g.}
\cite{mandel1,stern,angular,martinez2}).  Such an investigation is currently
underway by J.\ R.\ M..

\vskip 1cm
\noindent{\bf Acknowledgments} \\
Funding for this work provided by the Natural Sciences and Engineering
Research Council of Canada and the Walter C.\ Sumner Foundation.  Thanks
to Allen Attard for provision of the PSC libraries.

\include{PSCrefs}
\include{PSCtabsnfigs}

\end{document}

%% file: PSCtabsnfigs.tex
\begin{table}[h]
\begin{center}
{\begin{tabular}{l c c c c c}\hline
Survey & $D_F$ & Approx.\ Size \\ \hline
CfA1 & 1.7 (0.2)& 1800 \\
CfA2 & $\sim 2$& 11000\\
SSRS1 & 2.0 (0.1)& 1700\\
SSRS2 & $\sim 2$& 3600 \\
LEDA & 2.1 (0.2)& 75000\\
IRAS 1.2/2~Jy & 2.2 (0.2)& 5000 \\
Perseus-Pisces & $\sim 2.1$&3300  \\
ESP & 1.8 (0.2) & 3600 \\
Las Campa\~{n}as (LCRS) & 2.2 (0.2)& 25000\\ \hline
\end{tabular}}
\end{center}
\caption{
Galaxy fractal dimension calculations for various redshift surveys
(compiled from \cite{piet1}).
}
\label{galaxydims}
\end{table}

\pagebreak

\begin{table}[h]
\begin{center}
{\begin{tabular}{c c c c}\hline
Library (pts) & $D_0$ & $D_2$ & $D_{\infty}$\\ \hline
SCF1 (22076)& 2.5 (0.1) & 2.4 (0.1) & 2.2 (0.1)\\
%(22076)& 2.1 (0.2) & 2.2 (0.2) & 2.1 (0.2) & 2.1 (0.1)& 1.9 (0.1)\\ \hline
SCF2 (42325)& 2.6 (0.1) & 2.6 (0.1) & 2.2 (0.1) \\
%(42325)& 2.3 (0.2) & 2.3 (0.1) & 2.3 (0.1) & 2.2 (0.1) & 2.0 (0.2)\\ \hline
SCF3 (49797)& 2.7 (0.1) & 2.6 (0.1) & 2.2 ($<0.1$)\\
%(49797)& 2.4 (0.2) & 2.3 (0.1) & 2.3 (0.1) & 2.2 (0.1)& 2.0 (0.1)\\ \hline
SCF4 (73936)& 2.7 (0.1) & 2.6 (0.1) & 2.3 (0.1)\\
%(73936)& 2.5 (0.2) & 2.4 (0.1) & 2.4 (0.1) & 2.3 (0.1)& 2.1 (0.1)\\ \hline
SCF5 (82788)& 2.8 (0.1) & 2.6 (0.1) & 2.3 ($< 0.1$)\\ \hline
%(82788)& 2.5 (0.1) & 2.4 (0.1) & 2.4 (0.1) & 2.3 (0.1)& 2.2 (0.1)\\ \hline
\end{tabular}}
\end{center}
\caption{
Calculated dimensions for flat space packings.  Cube size ranges from
homogeneous saturation at largest scales (8 cubes) to roughly 2-4 points
per cube (ranging from scales ($\sim \pi/10 - \pi/300$)).
}
\label{flat1}
\end{table}

\pagebreak

\begin{table}[h]
\begin{center}
{\begin{tabular}{c c c c}\hline
Library (pts)& $D_0$ & $D_2$ & $D_{\infty}$ \\ \hline
SCP1 (31904)& 2.6 (0.1) & 2.5 (0.1) & 2.2 (0.1) \\
%(31904) & 2.3 (0.2) & 2.3 (0.1) & 2.2 (0.1) & 2.1 (0.1) & 1.8 (0.2) \\ \hline
SCP2 (35268)& 2.7 (0.1) & 2.6 (0.1) & 2.2 (0.1)\\
%(35268) & 2.4 (0.2) & 2.3 (0.2) & 2.2 (0.1) & 2.1 (0.2) & 1.9 (0.2)\\ \hline
SCP3 (47140)& 2.7 (0.1) & 2.6 (0.1) & 2.2 (0.1) \\
%(47140) & 2.4 (0.1) & 2.4 (0.1) & 2.3 (0.1) & 2.2 (0.1) & 1.9 (0.1) \\ \hline
SCP4 (54966)& 2.7 (0.1) & 2.6 (0.1) & 2.2 (0.1) \\
%(54966) & 2.4 (0.1) & 2.4 (0.1) & 2.3 (0.1) & 2.2 (0.1) & 1.7 (0.2)  \\ \hline
SCP5 (80437)& 2.8 (0.1) & 2.6 (0.1) & 2.3 (0.1) \\  \hline
%(80437) & 2.5 (0.1) & 2.5 (0.1) & 2.4 (0.1) & 2.3 (0.1) & 2.1 (0.1) \\ \hline
\end{tabular}}
\end{center}
\caption{
Calculated dimensions for positively-curved space packings.
}
\label{pos1}
\end{table}

\pagebreak

\begin{table}[h]
\begin{center}
{\begin{tabular}{c c c c}\hline
Library & $D_0$ & $D_2$ & $D_{\infty}$ \\ \hline
SCN1 (32965)& 2.6 (0.1) & 2.5 (0.1) & 2.1 (0.1) \\
%(32965) & 2.4 (0.2) & 2.4 (0.1) & 2.3 (0.1) & 2.3 (0.1) & 1.9 (0.2) \\ \hline
SCN2 (49343)& 2.7 (0.1) & 2.6 (0.1) & 2.2 (0.1) \\
%(49343) & 2.4 (0.2) & 2.3 (0.1) & 2.3 (0.1) & 2.2 (0.1) & 2.0 (0.1) \\ \hline
SCN3 (60245)& 2.7 (0.1) & 2.6 (0.1) & 2.3 (0.1) \\
%(60245) & 2.4 (0.2) & 2.4 (0.1) & 2.3 (0.1) & 2.3 (0.1) & 2.1 (0.1) \\ \hline
SCN4 (65899)& 2.7 (0.1) & 2.6 (0.1) & 2.3 (0.1) \\
%(65899) & 2.5 (0.1) & 2.4 (0.1) & 2.3 (0.1) & 2.3 (0.1) & 2.1 (0.1) \\ \hline
SCN5 (83863)& 2.8 (0.1) & 2.6 (0.1) & 2.3 (0.1) \\ \hline
%(83863) & 2.5 (0.1) & 2.5 (0.1) & 2.4 (0.1) & 2.3 (0.1) & 2.1 (0.2)\\ \hline
\end{tabular}}
\end{center}
\caption{
Calculated dimensions for negatively-curved space packings.
}
\label{neg1}
\end{table}

\pagebreak

\begin{table}[h]
\begin{center}
{\begin{tabular}{l l c c c c}\hline
Lib. & \# (\% tot) & \% mass & $D_0$ & $D_2$ & $D_{\infty}$ \\ \hline
SCF1 & 4526 (20.5) & 96.0 & 2.1 (0.3) & 2.1 (0.2) & 1.9 (0.1) \\
SCF2 & 5074 (12.0) & 93.6 & 2.1 (0.3) & 2.1 (0.2) & 1.9 (0.1) \\
SCF3 & 5218 (10.5) & 92.8 & 2.2 (0.3) & 2.2 (0.2) & 1.9 (0.1) \\
SCF4 & 5506 (7.4) & 91.1 & 2.2 (0.3) & 2.2 (0.2) & 1.9 (0.1) \\
SCF5 & 5599 (6.7) & 90.6 & 2.2 (0.3) & 2.2 (0.2) & 1.9 (0.1) \\ \hline
SCP1 & 4744 (14.9) & 94.6 & 2.1 (0.3) & 2.1 (0.2) & 1.8 (0.2) \\
SCP2 & 5848 (16.6) & 93.3 & 2.3 (0.3) & 2.2 (0.2) & 2.0 (0.1)\\
SCP3 & 5108 (10.8) & 92.9 & 2.2 (0.3) & 2.2 (0.2) & 2.0 (0.1)\\
SCP4 & 5233 (9.5) & 92.4 & 2.2 (0.3) & 2.2 (0.2) & 1.9 (0.1) \\
SCP5 & 5542 (6.9) & 90.6 & 2.2 (0.3) & 2.2 (0.2) & 1.8 (0.2) \\ \hline
SCN1 & 4993 (15.1) & 94.7 & 2.1 (0.3) & 2.1 (0.2) & 1.9 (0.1)\\
SCN2 & 5294 (10.7) & 93.0 & 2.2 (0.3) & 2.2 (0.2) & 1.9 (0.1) \\
SCN3 & 5409 (9.0) & 92.1 & 2.2 (0.3) & 2.2 (0.2) & 1.9 (0.1)\\
SCN4 & 5536 (8.4) & 91.7 & 2.2 (0.3) & 2.2 (0.2) & 1.9 (0.1)\\
SCN5 & 5837 (7.0) & 90.6 & 2.2 (0.3) & 2.2 (0.2) & 2.0 (0.2)\\ \hline
\end{tabular}}
\end{center}
\caption{
Mass-reduced box counts for all libraries, showing box, correlation, and
limiting $D_{q \rightarrow \infty}$ values.  Lower $D_0$ values with higher
errors may be artificially lower due to finiteness effects.
}
\label{allbias}
\end{table}

\pagebreak

\begin{figure}[h] \begin{center} \leavevmode
\includegraphics[width=0.6\textwidth]{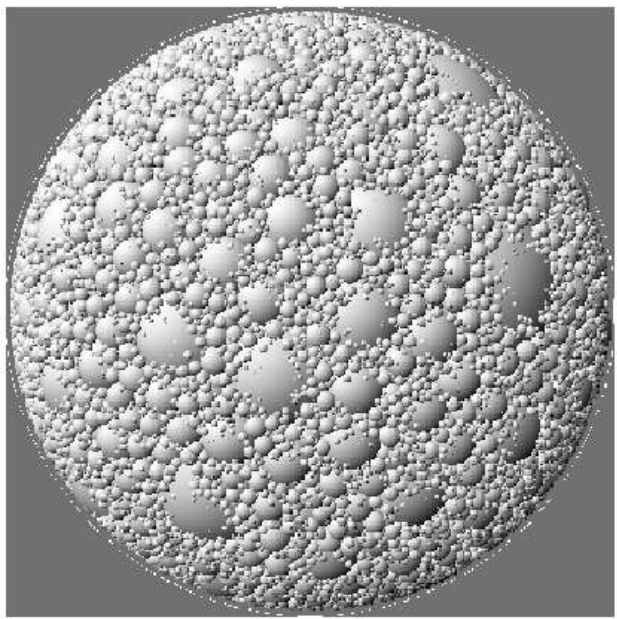}
\end{center} \caption{
Sample projected two-dimensional visualization of packing library, containing
approximately 35000 spheres.  Surfaces represent the inscribed radius, and
not the actual contained matter.
}
\label{packgif}
\end{figure}

\pagebreak

\begin{figure}[h] \begin{center} \leavevmode
\includegraphics[angle=270,width=1.0\textwidth]{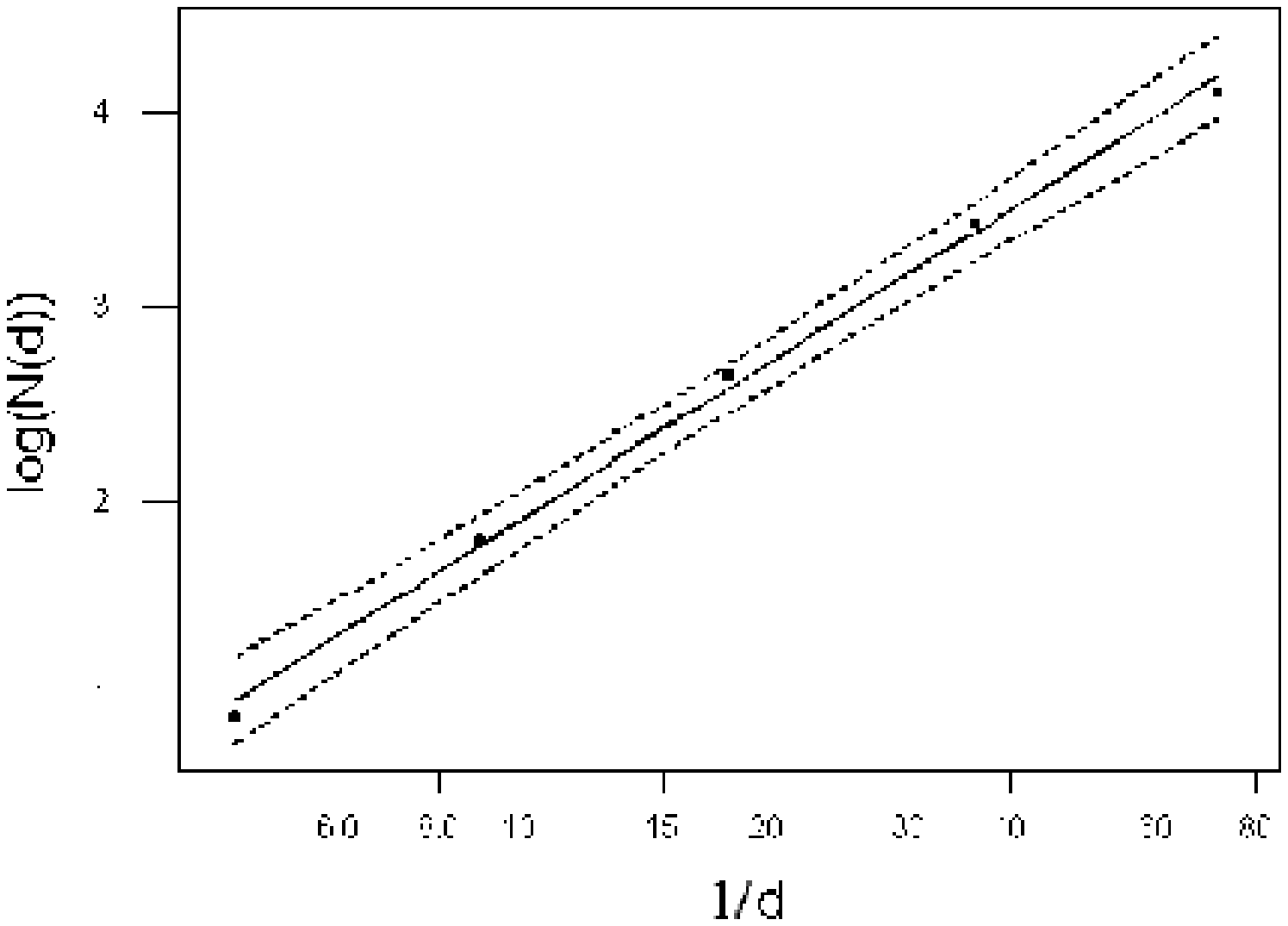}
\end{center} \caption{
Box counting ($D_0$) fit for SCF3, yielding $D_0 = 2.7~(0.1)$.
}
\label{flatfit}
\end{figure}
\begin{figure}[h] \begin{center} \leavevmode
\includegraphics[angle=270,width=1.0\textwidth]{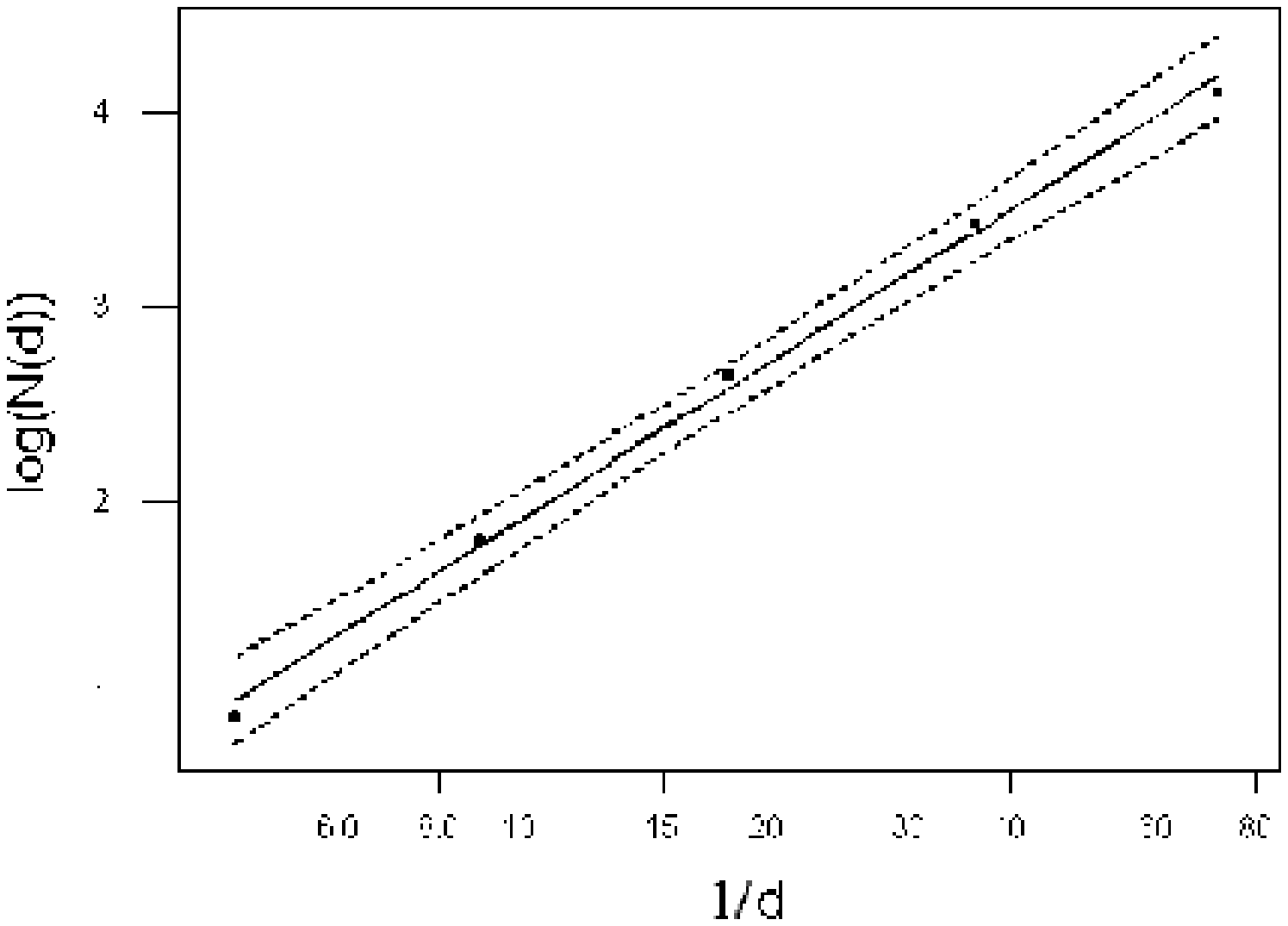}
\end{center} \caption{
Box counting ($D_0$) fit for SCP3 yielding $D_0 = 2.7~(0.1)$.
}
\label{posfit}
\end{figure}
\begin{figure} \begin{center} \leavevmode
\includegraphics[angle=270,width=1.00\textwidth]{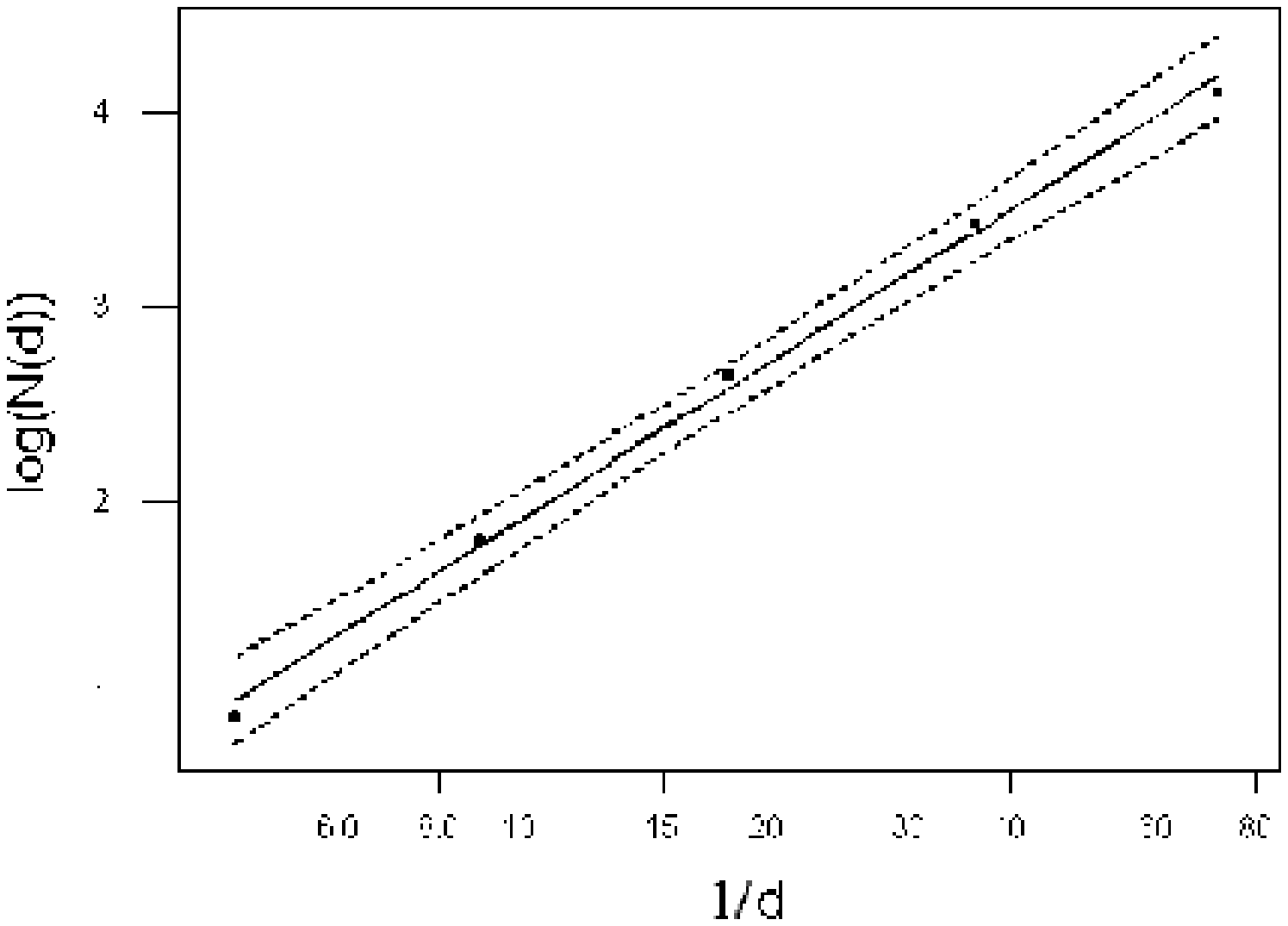}
\end{center} \caption{
Box counting ($D_0$) fit for SCN3  yielding $D_0 = 2.7~(0.1)$.
}
\label{negfit}
\end{figure}

\pagebreak

\begin{figure}[h] \begin{center} \leavevmode
\includegraphics[angle=270,width=0.7\textwidth]{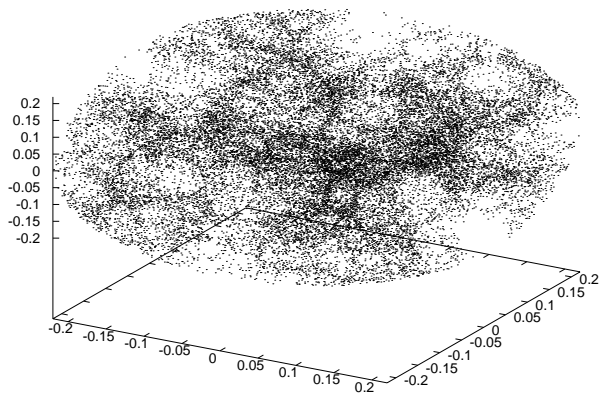} \vskip .5cm
\includegraphics[angle=270,width=0.7\textwidth]{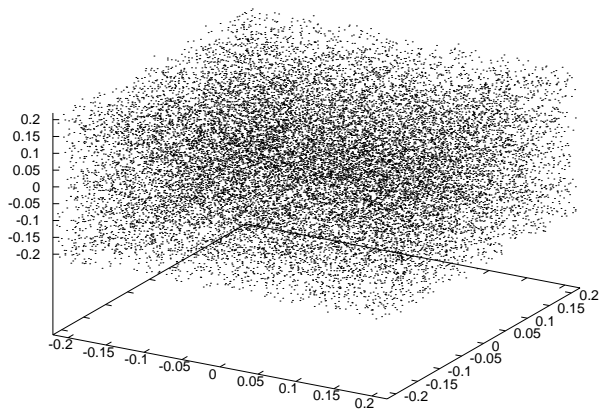}
\end{center} \caption{
Three-dimensional point distributions, including: packing library
(SCF3), randomly-populated $(x,y,z)$ coordinates (RD).
Each set possesses between 50000-60000 points.
}
\label{randsets1}
\end{figure}
\begin{figure}[h] \begin{center} \leavevmode
\includegraphics[angle=270,width=0.7\textwidth]{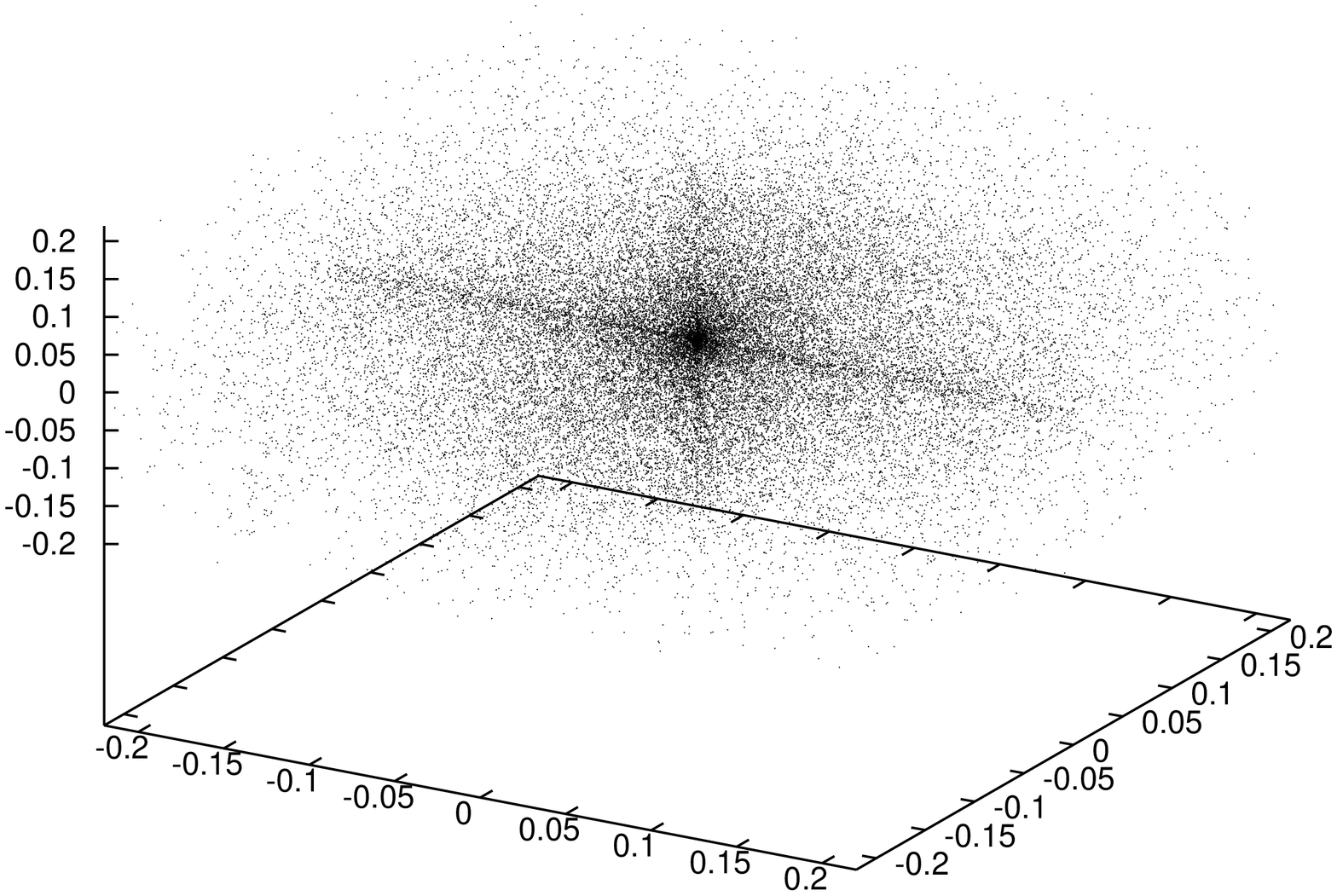} \vskip .5cm
\includegraphics[angle=270,width=0.7\textwidth]{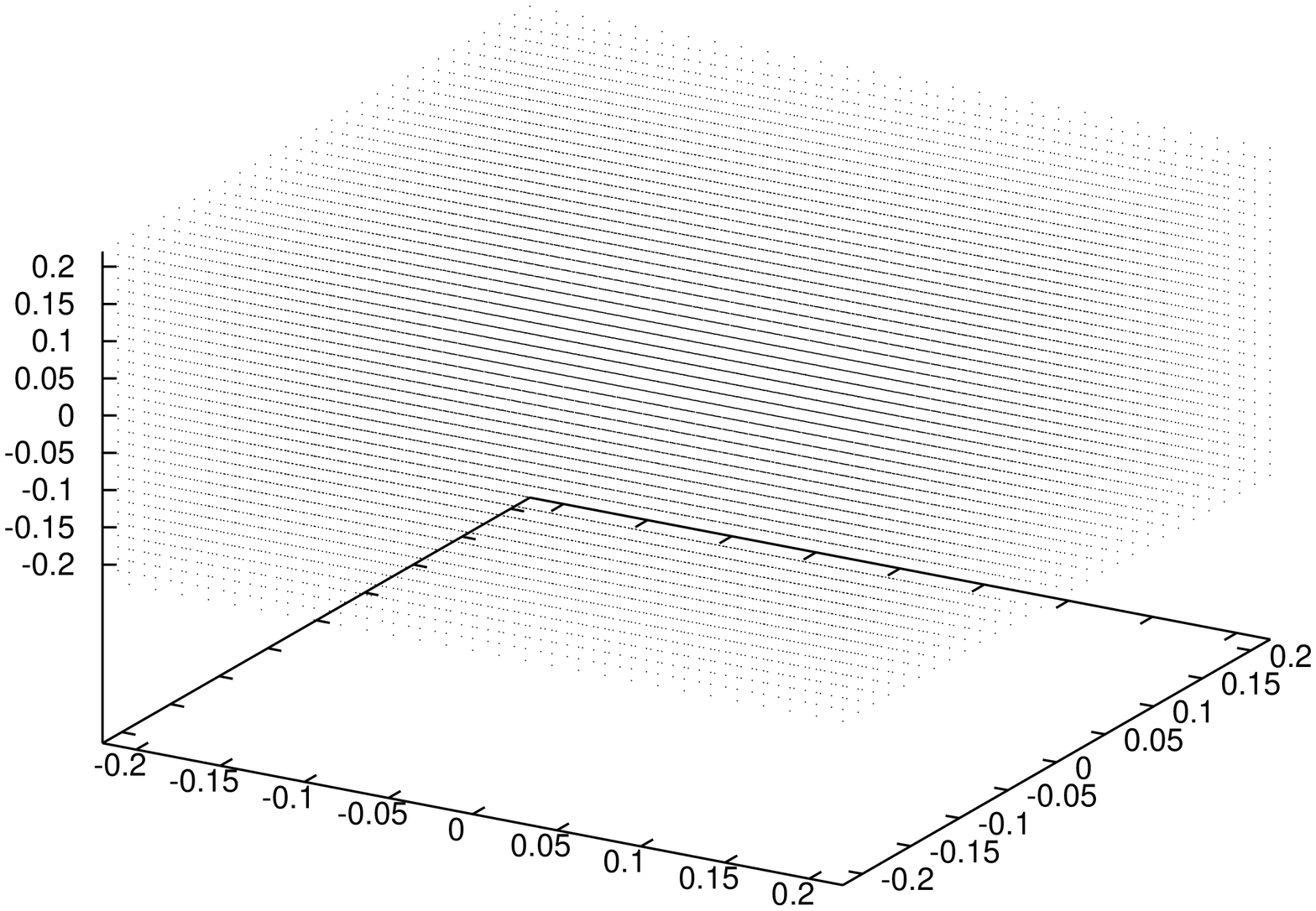}
\end{center} \caption{
Three-dimensional point distributions, including
linear-random (LR; top), and uniform distribution (UD; bottom).
Each set possesses between 50000-60000 points.  The LR
set is comprised of randomly-populated $(r,\theta,\phi)$ points,
in contrast to the random population of $(x,y,z)$ coordinates.
}
\label{randsets2}
\end{figure}

\pagebreak

\begin{figure}[h] \begin{center} \leavevmode
\includegraphics[angle=270,width=0.75\textwidth]{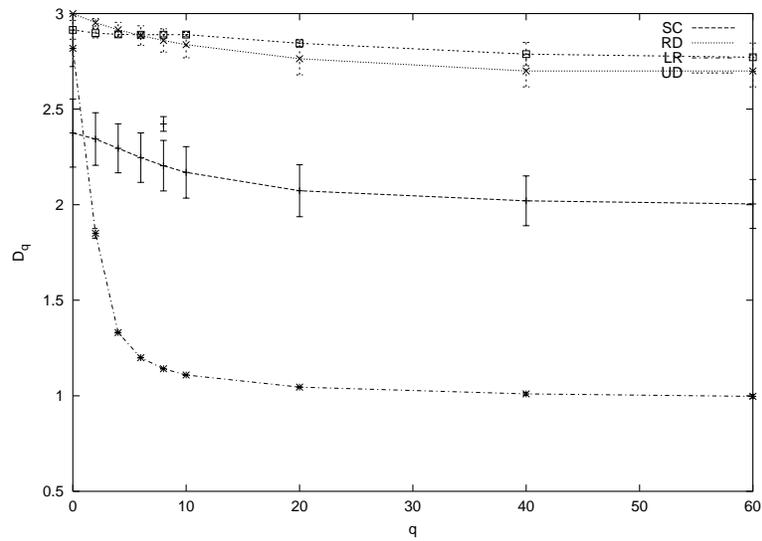}
\end{center} \caption{
$D_q$ spectra ($q > 0$) for distributions of
Figure~\ref{randsets1},\ref{randsets2}.  All sets show high $D_0$ values,
with the RD and UD sets (top two curves) maintaining $D \sim 3$.
Explicit structural differences can be seen in the SC and LR sets
(bottom two curves) for large $q$, signifying strong inhomogeneous
clustering behaviors.
}
\label{setdqs}
\end{figure}

\pagebreak

\begin{figure}[h] \begin{center} \leavevmode
\includegraphics[angle=270,width=0.75\textwidth]{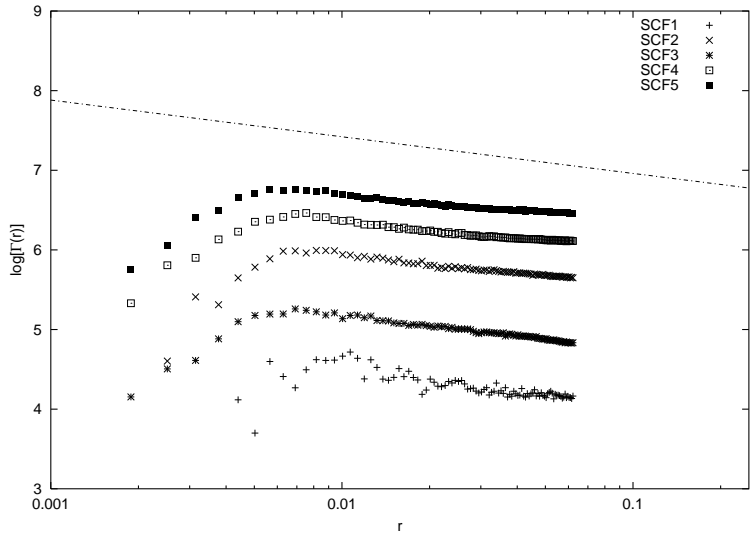}
\end{center} \caption{
Conditional density $\Gamma(r)$ for flat
libraries.  Sample
fit (dashed line) has slope $D-3 = -0.2$~, which implies $D = 2.8$.
Note that each library has been shifted vertically to eliminate
overlap for easier viewing.
}
\label{cdensfig1}
\end{figure}
\begin{figure}[h] \begin{center} \leavevmode
\includegraphics[angle=270,width=0.75\textwidth]{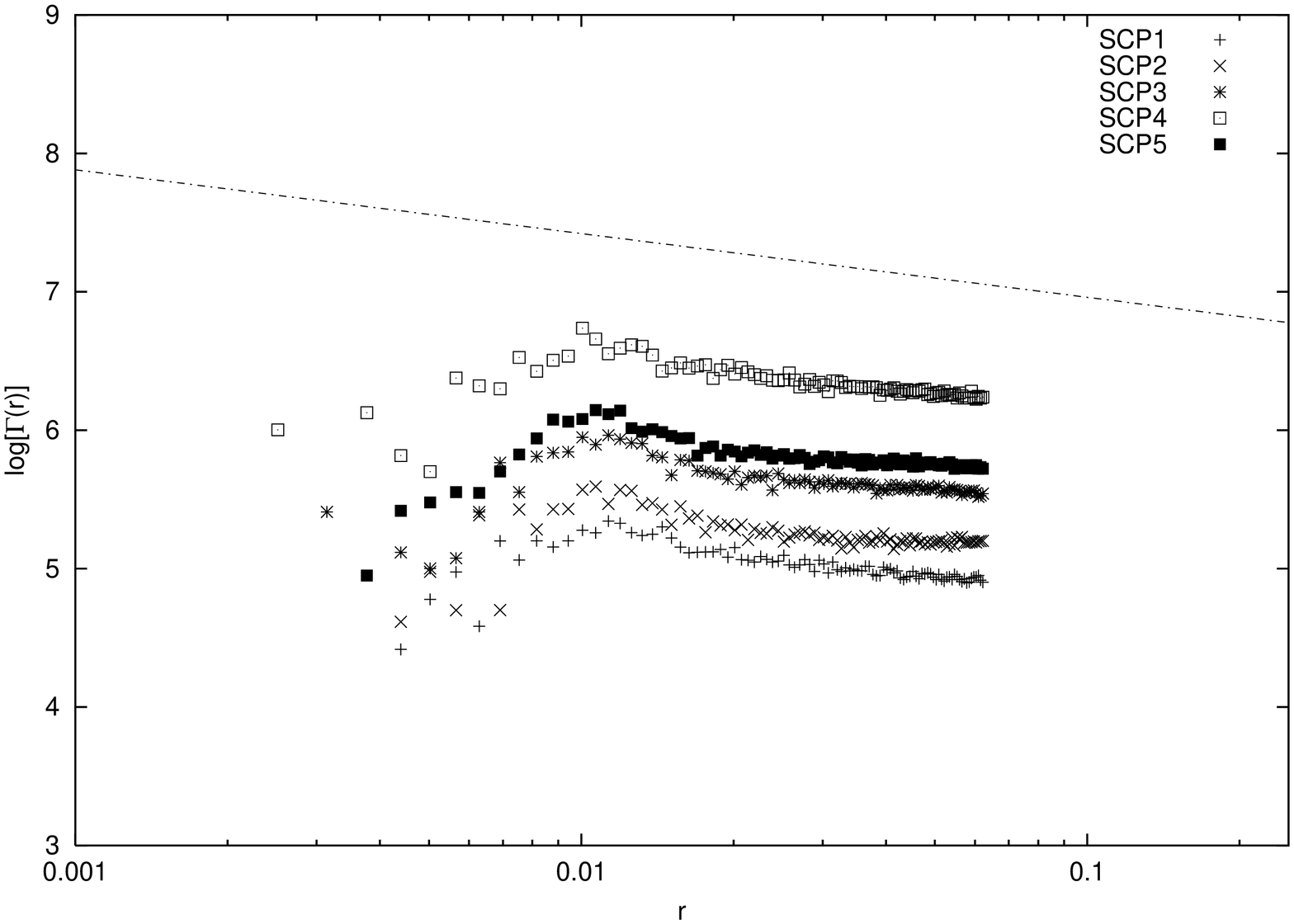}
\end{center} \caption{
Conditional density $\Gamma(r)$ for positive
libraries.  Sample
fit (dashed line) has slope $D-3 = -0.2$~, which implies $D = 2.8$.
}
\label{cdensfig2}
\end{figure}
\begin{figure}[h] \begin{center} \leavevmode
\includegraphics[angle=270,width=0.75\textwidth]{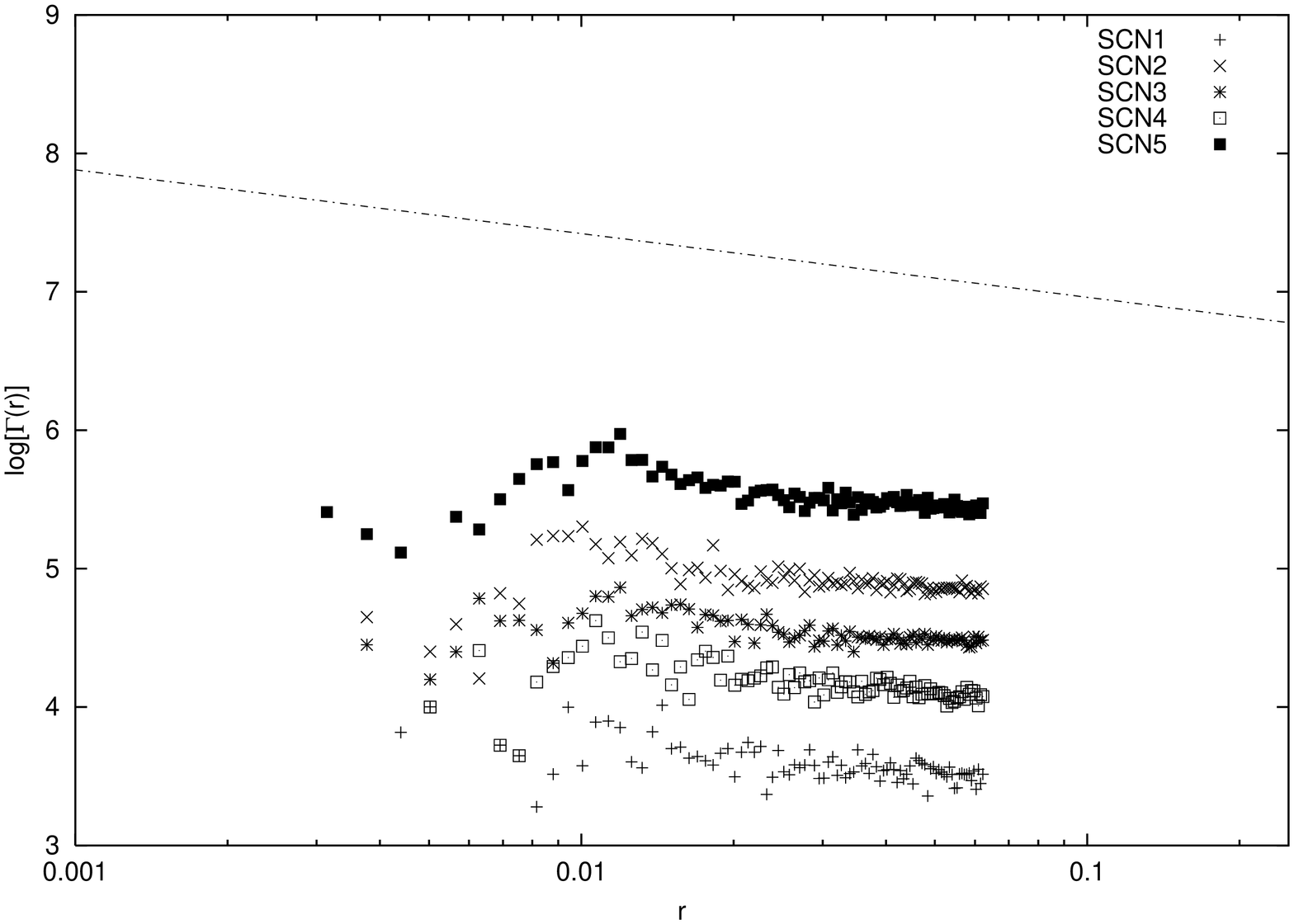}
\end{center} \caption{
Conditional density $\Gamma(r)$ for
negative libraries.  Sample
fit (dashed line) has slope $D-3 = -0.2$~, which implies $D = 2.8$.
}
\label{cdensfig3}
\end{figure}

\pagebreak

\begin{figure}[h] \begin{center} \leavevmode
\includegraphics[angle=270,width=0.75\textwidth]{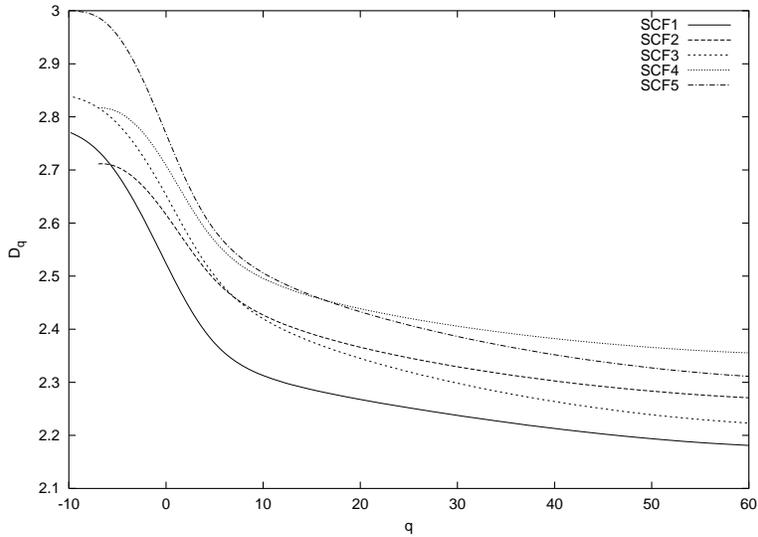}
\end{center} \caption{
Bezier curve fit of full $D_q$ spectra for flat
packing libraries, showing
$D_q \rightarrow \sim 3$ for small $q$.  $q < 0$
values calculated by density reconstruction; $q \geq 0$ by box counting.
}
\label{dqlibs1}
\end{figure}
\begin{figure}[h] \begin{center} \leavevmode
\includegraphics[angle=270,width=0.75\textwidth]{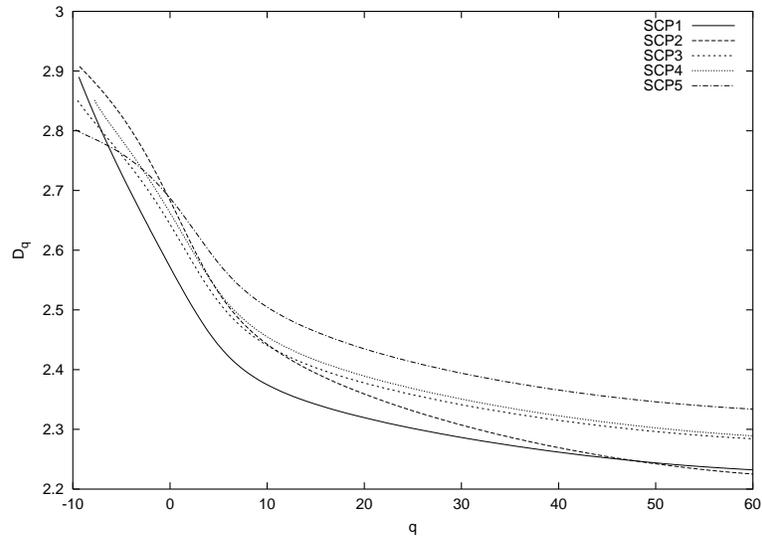}
\end{center} \caption{
Bezier curve fit of full $D_q$ spectra for positive
packing libraries.
}
\label{dqlibs2}
\end{figure}

\begin{figure}[h] \begin{center} \leavevmode
\includegraphics[angle=270,width=0.75\textwidth]{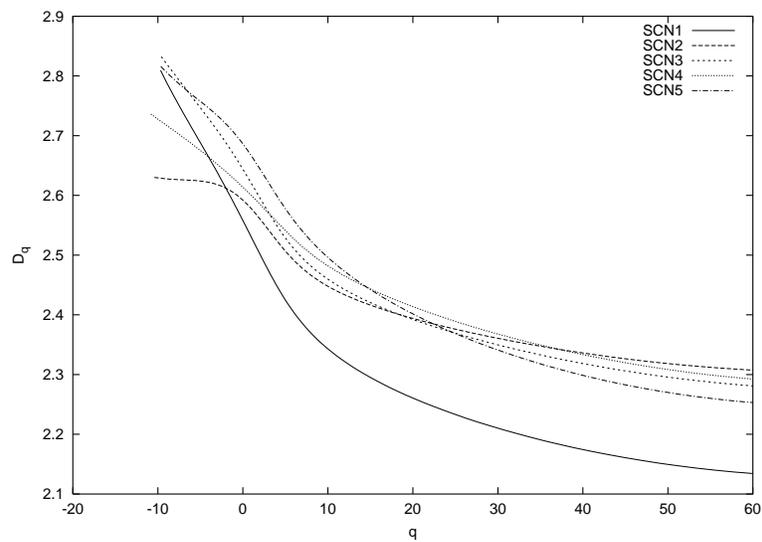}
\end{center} \caption{
Bezier curve fit of full $D_q$ spectra for
negative curvature packing libraries.
}
\label{dqlibs3}
\end{figure}

\pagebreak

\begin{figure}[h] \begin{center} \leavevmode
\includegraphics[angle=270,width=1.0\textwidth]{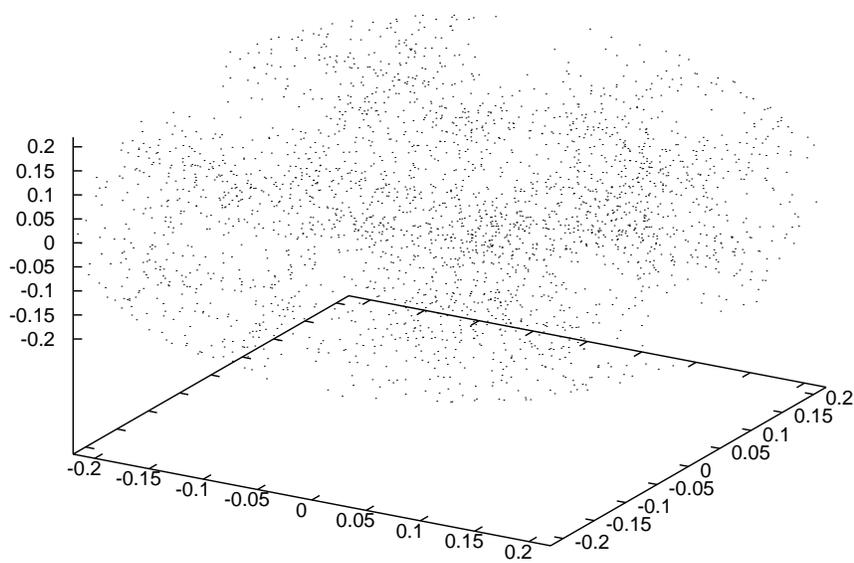}
\end{center} \caption{
Library SCF3 with effective mass cutoff, containing 10\% of spheres representing
approximately 90\% of total contained mass.
}
\label{pac10cut1}
\end{figure}

\begin{figure}[h] \begin{center} \leavevmode
\includegraphics[angle=270,width=0.75\textwidth]{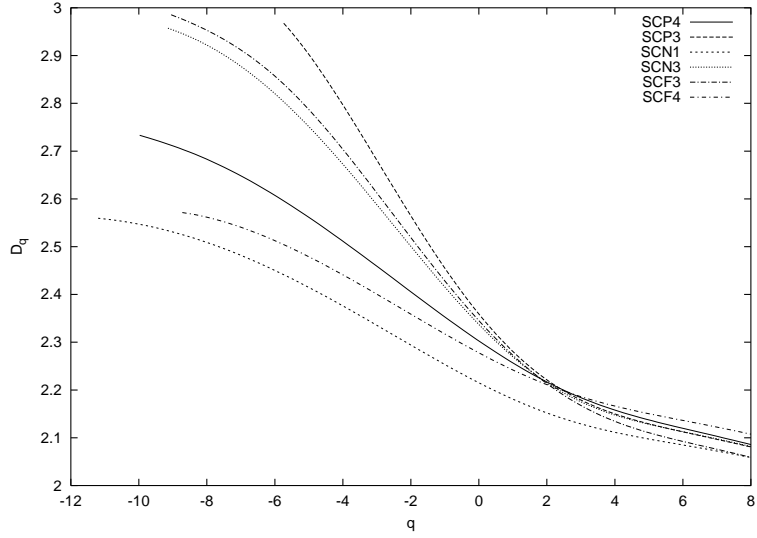}
\end{center} \caption{
Bezier curve fit of full $D_q$ spectra for mass-reduced libraries, showing
$D_q \rightarrow \sim 3$ for small $q$.  $q < 0$
values calculated by density reconstruction; $q \geq 0$ by box counting.
}
\label{dqnegs}
\end{figure}

\pagebreak

\begin{figure}[h] \begin{center} \leavevmode
\includegraphics[angle=270,width=0.75\textwidth]{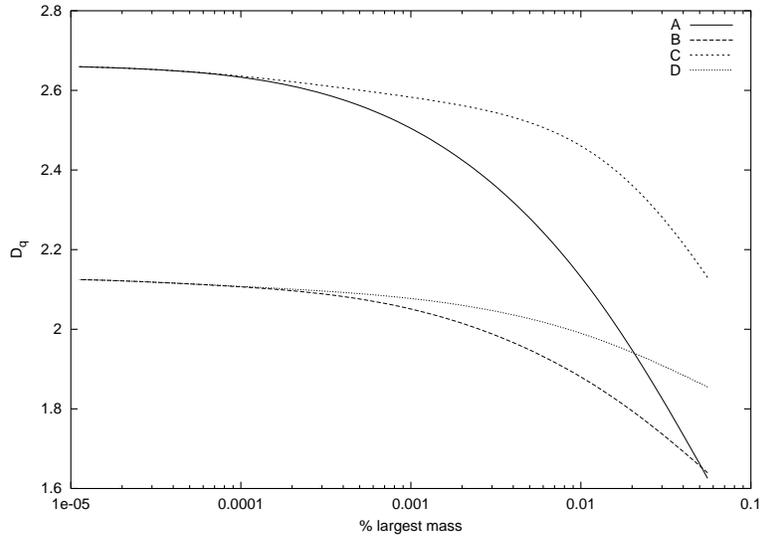}
\end{center} \caption{
Variation of $D_0$ (A, C) and $D_{\infty}$ (B, D)
as a function of cutoff (\% of largest mass) for fixed number of box level
iterations (A, B) and ``last-point-removed'' (C, D).  True fractal
dimension is bounded by the two curves.
}
\label{mcvar}
\end{figure}